# Unconditional Security in Quantum Cryptography

DOMINIC MAYERS

*NEC Research Institute, Princeton, New Jersey*

Abstract. Basic techniques to prove the unconditional security of quantum cryptography are described. They are applied to a quantum key distribution protocol proposed by Bennett and Brassard [1984]. The proof considers a practical variation on the protocol in which the channel is noisy and photons may be lost during the transmission. Each individual signal sent into the channel must contain a single photon or any two-dimensional system in the exact state described in the protocol. No restriction is imposed on the detector used at the receiving side of the channel, except that whether or not the received system is detected must be independent of the basis used to measure this system.

## 1. Introduction

This paper proves the unconditional security of quantum key distribution and reviews basic notions and principles which apply to any quantum key distribution protocol, and in fact to other kind of quantum protocols as well. The protocol that we consider was proposed by Bennett and Brassard [1984], which was also the first proposed quantum key distribution protocol. An improved variation on the protocol was proposed later in Bennett et al. [1992]. A first version of the proof was published in Mayers [1996]. The proof relies on techniques provided in Mayers and Salvail [1994], Yao [1995], and Mayers [1995], but the paper is self-contained.

At the time of writing, all other known proofs of security in quantum cryptography consider only restricted kinds of attacks.[1] Though some of these

---

[1] See, for example, Bennett and Brassard [1984], Bennett et al. [1966; 1992], Mayers and Salvail [1994], Ekert [1991], Bennett [1992], Deutsch et al. [1996], Biham and Mor [1996], and Biham et al. [1998].

---

This research was partially supported by DIMACS and part of the work was done while the author worked for the Department of Computer Science of Princeton University.

Author's address: D. Mayers, NEC Research Institute, 4 Independence Way, Princeton, NJ 08540; e-mail: mayers@research.nj.nec.com.

other proofs[2] encompass all attacks that can be realized with the current technology, it is of interest to establish a security guarantee that holds against unlimited computational power, both classical and quantum mechanical.

In Deutsch et al. [1996], the authors discuss how security (for a different kind of quantum key distribution protocols [Ekert 1991; Bennett et al. 1996]) against all attacks can be obtained. However, additional work was required to obtain a bound on the eavesdropper's information that is valid for all attacks or to obtain a proof that such a bound exists (see the security criteria in Section 2). A bound was proposed recently under the assumption that the honest participants use perfect quantum computers and quantum communication devices [Lo and Chau 1998]. The above list of references is not intended to be exhaustive. A survey of previous works in quantum cryptography can be found in Brassard and Crépeau [1996].

In quantum key distribution, and ideally in other applications of quantum cryptography, a security result is expected to hold against all attacks allowed by quantum mechanics. This is what is called an *unconditional security*, and this is what we will prove. On top of an unconditional security, it is desirable to have a security that holds when quantum computing devices (such as measuring apparatus, sources of photons, or quantum channels) used in the protocol are imperfect. A fundamental aspect of security is to know where the trust is required. The property that needs to be trusted should be reasonable and simple. Here, we will take care of an imperfect measuring apparatus and an imperfect channel in the protocol analyzed. However, we will assume that the source transmits a single photon per pulse with the exact polarization angle specified in the protocol. Putting our trust in this property is not so reasonable, but fortunately a fundamental mechanism that can take care of this issue is already known [Mayers and Yao 1998; Mayers 2001a].

Section 2 defines the general notion of privacy for quantum key distribution. Section 3 contains preliminaries, basic lemmas and the general model used to analyze the protocol. In section 4, the protocol is described. Section 5 contains the proof of privacy. The notation used is summarized in Appendix A.

## 2. *The Security Criteria*

In this article, a security definition that applies to all quantum key distribution protocols is provided. The goal of an ideal key distribution is to allow two participants, Alice and Bob, who share no information initially to share a secret key (a string of bits) at the end. A third participant, usually called Eve, should not be able to obtain any information about the key. Also, whatever Eve does, Alice's and Bob's key should be identical. It is assumed that all quantum communication between Alice and Bob passes through Eve, and similarly for classical communication.

2.1. THEORETICAL IMPOSSIBILITIES IN KEY DISTRIBUTION.   In reality, one cannot realize this ideal task. There are few subtle points to consider. In particular, no quantum key distribution protocol can succeed if Eve has the power to impersonate Alice while communicating with Bob and to impersonate Bob while communicating with Alice. To address this problem, Alice and Bob can authenticate [Wegman and Carter 1981] their classical messages so that Eve cannot impersonate them any more. There exist unconditionally secure techniques for authentication [Wegman and Carter 1981] that require that Alice and Bob share a small secret key to begin with, so that the protocol implements key *expansion* rather than key distribution [Bennett et al. 1992]. This approach can be used in a scenario where Alice and Bob have met before to exchange the initial key. In a scenario where Alice and Bob have never exchanged a secret key before, one must assume that Alice and Bob have access to a faithful (classical) public channel so that a third party cannot accomplish the impersonation attack without being detected. The cheater can only add his messages to the messages which are faithfully exchanged between Alice and Bob, but then Alice and Bob will see that a third person is trying to cheat. This principle was previously mentioned in Bennett et al. [1992].

---

[2] See, for example, Mayers and Salvail [1994], Bennett et al. [1966], Deutsch et al. [1996], Biham and Mor [1996], and Biham et al. [1998].

Another related point is that a secret key is not always shared between Alice and Bob because it is always possible for a third party to jam the quantum channel. Alice and Bob must verify that some validation constraints are satisfied, including an upper bound on the number of errors, and decide accordingly whether or not they can share a secret key. These validation constraints encompasses anything that Alice and Bob consider in order to decide whether or not they can share a secret key. The event that the validation constraints are satisfied is denoted $\mathscr{P}$.

2.2. ON SECURITY PARAMETERS. We typically define the security of a protocol in terms of a security parameter $N$. In an information theoretic setting, which is our case, a quantity $f_N$ such as the amount of Shannon's information available to Eve must decrease exponentially fast as $N$ increases. Often, there are other parameters in the protocol, for example, the tolerated error rate or the percentage of bits used for a test. Let $\bar{\epsilon}$ denotes all these parameters. The value of some of the parameters in $\bar{\epsilon}$, such as the error rate, depends upon the physical setting. The other parameters are not fixed by the physical setting. Ideally, we should optimize the protocol, that is, minimize $f_N$, over the nonphysical parameters. In any case, we want to prove that the quantity $f_N = f_N(\bar{\epsilon})$ is exponentially small in $N$ for all valid values of these parameters. The implicit quantifiers are ($\forall \bar{\epsilon} \in$ ValidDomain) ($\forall N \geq N_0(\bar{\epsilon})$) $f_N(\bar{\epsilon}) \leq c(\bar{\epsilon}) \exp(-g(\bar{\epsilon})N)$ where $N_0$, $c$ and $g$ are well-defined functions on $\bar{\epsilon} \in$ ValidDomain. In an information theoretic setting, we believe that the above is the essential that one needs to know about security parameters. The situation is more complicated in a computational setting.

We can do the complete analysis of a protocol for fixed values of $N$ and $\bar{\epsilon}$. If these were the actual values used in a given application, such an analysis will be enough for this particular application. Since the use of security parameters in statements that makes sense without them will unnecessarily complicate the notation, we will not use security parameters in our basic definitions and general results. However, the statements about the specific protocol that we analyze will contain security parameters. To obtain the asymptotic behavior, we will simply make sure that the proof respects the quantifiers ($\forall \bar{\epsilon} \in$ ValidDomain) ($\forall N \geq N_0(\bar{\epsilon})$)..., but that is an easy mathematical task. The important concepts apply to fixed values of the parameter $N$ and $\bar{\epsilon}$.

2.3. On Random Variables.   Formally, a *random variable* **x** is a function on a probabilistic space, the space of every possible outcome of a random experiment. In our case, this probabilistic space corresponds to the set of every possible outcome $\hat{v}$ of the protocol. The outcome $\hat{v}$ includes all data generated in the protocol. Its probability is denoted $p(\hat{v})$. So, a random variable $\mathbf{x} = \mathbf{x}(\hat{v})$ takes a value $x$ which depends upon the outcome $\hat{v}$. It is the standard convention for random variables to write $\mathbf{x} = x$ to mean $\mathbf{x}(\hat{v}) = x$, that is, we do not explicitly write the dependence upon the outcome of the random experiment. Usually, we will use a boldface typesetting for a random variable, that is, the function, when it is important to distinguish it from its value. Variables are also used to describe protocols and procedures. These are other kind of variables, and we will use ordinary typesetting in this case.

In this paper, an event is a random variable with outcome True or False, that is, it is a property of the random outcome $\hat{v}$ of the protocol. In other papers, an event is defined as a set of outcomes. This alternative definition is equivalent. The probability of an event $\mathscr{E}$ is $\Pr(\mathscr{E}) \stackrel{def}{=} \Sigma_{\hat{v}|\mathscr{E}}\, p(\hat{v})$.

If for every element $\hat{v}$ in the probability space, we have $\mathbf{y}(\hat{v}) = f(\mathbf{x}(\hat{v}))$ for some deterministic function $f$, we say that the random variable **y** is a deterministic function of **x**, and we write $\mathbf{y}_{|\mathbf{x}}(x)$ to denote the value taken by **y** when $\mathbf{x} = x$. We will omit the indice "$|\mathbf{x}$" when the situation permits.

2.4. The Criteria.   To formulate our criteria, we will consider the protocol as a random experiment that defines random variables. Let *Eve's view* **v** be all the classical data received or generated by Eve during the protocol. This includes classical announcements and outcomes of measurements. Here, it is assumed that the result of the validation test is announced to Eve at the end of the protocol, so that the event $\mathscr{P}$ (True when the test succeeds) is a deterministic function of **v**. We often associate security of key distribution with privacy, but the security of key distribution also includes that Alice's and Bob's keys must be identical. We shall consider this other security aspect as well, but we understand that the interesting aspect is privacy.

For privacy, it is assumed that Eve is interested about Alice's key, which is denoted **k**. The length of the key is always defined and included in Eve's view **v**. The length of the key does not have to be fixed in advance, and it may be convenient not to fix it. For example, to use the channel at its full capacity which may vary in time (in view of the weather, etc.), the length of the key can be made a function of the error rate measured during the execution of the protocol. When the test fails, the protocol sets $\mathbf{m} = 0$ and **k** is the null string.

For any two random variables **x** and **y**, we denote by $p_{\mathbf{x}}(x) = \Pr(\mathbf{x} = x)$ and $p_{\mathbf{x}|\mathbf{y}}(x|y) = \Pr(\mathbf{x} = x|\mathbf{y} = y)$. However, when the situation permits, we will use $p(x)$, $p(x|y)$, etc, to mean $p_{\mathbf{x}}(x)$, $p_{\mathbf{x}|\mathbf{y}}(x|y)$, etc.

*Definition* 1.   Consider any number $f > 0$. A quantum key distribution protocol is *f*-private if, for every strategy adopted by Eve,

$$\sum_m p(m)(m - H_m(\mathbf{k}|\mathbf{v})) \le f \qquad (1)$$

*where*

$$H_m(\mathbf{k}|\mathbf{v}) \overset{def}{=} -\sum_{v} \sum_{k \in \{0,1\}^m} p(k, v|m) \log_2(p(k|v))$$

is the Shannon entropy of the key $\mathbf{k}$ conditional to Eve's view $\mathbf{v}$ in the context of a fixed length $m$ for the key.

One would think that $I(\mathbf{k};\mathbf{v}) \leq f$, where $I(\mathbf{k};\mathbf{v})$ is the mutual information (see Appendix B), should be the privacy criteria. To the contrary, our definition says that the privacy criteria is simply that the key must be uniformly distributed. The average of the quantity $m - H_m(\mathbf{k}|\mathbf{v})$, not the mutual information $I(\mathbf{k};\mathbf{v})$, corresponds to information that Eve has about the key. The idea is that Eve's attack might influence the distribution of the key $\mathbf{k}$ independently of her view $\mathbf{v}$, that is, Eve's attack might even influence the *a priori* distribution of probability of the key $\mathbf{k}$. In an unrealistic example, Eve might attack the protocol in such a way that the only possible keys are the keys $k = 0 \cdots 0$ with 0 everywhere. Given this attack, even before the protocol runs, she knows that the key will be $0 \cdots 0$. A distinction between the information that is available a priori (i.e., before Eve receives any data) and the information that is obtained a posteriori (i.e., via Eve's view $\mathbf{v}$) seems unnecessary and artificial.

Note that in the definition of $H_m(\mathbf{k}|\mathbf{v})$ one can think that the sum runs over the values of $v$ such that $\mathbf{m}(v) = m$ because $p(k, v|m) = 0$ when $\mathbf{m}(v) \neq m$. When $\mathbf{m}(v) = m$, we have $p(k, v, m) = p(k, v)$ and $p(k, v|m) = p(k, v)/p(m)$. Therefore, we have

$$H_m(\mathbf{k}|\mathbf{v}) \overset{def}{=} -\sum_{v|\mathbf{m} = m} \sum_{k \in \{0,1\}^m} \frac{p(k, v)}{p(m)} \log_2(p(k|v))$$

We obtain

$$\sum_m p(m) H_m(\mathbf{k}|\mathbf{v}) = -\sum_{k,v} p(k, v) \log_2 p(k|v)$$

where the length of the key $k$ in the right hand side runs over all nonnegative integers. Similarly, using $p(m) = \Sigma_{k,v|\mathbf{m}(v)=m} p(k, v)$, we have

$$\sum_m p(m)m = \sum_m \left[ \sum_{k,v|\mathbf{m}(v) = m} p(k, v) \right] m = \sum_{k,v} p(k, v)\mathbf{m}(v),$$

where, again, the length of the key $k$ in the right-hand side runs over all nonnegative integers. We obtain

$$\sum_m p(m)(m - H_m(\mathbf{k}|\mathbf{v})) = \sum_{k,v} p(k, v)(\mathbf{m}(v) + \log_2 p(k|v)), \qquad (2)$$

which can be used to reformulate (1). This can be used to check that, if $p(k|\mathbf{m}(v) = m) = 2^{-m}$ for every $k$ and $m$, then

$$\sum_m p(m)(m - H_m(\mathbf{k}|\mathbf{v})) = I(\mathbf{k};\mathbf{v}|\mathbf{m}),$$

where

$$I(\mathbf{k};\mathbf{v}|\mathbf{m}) = \sum_{k,v} p(k,v)\log_2\left(\frac{p(k,v|m(v))}{p(k|m(v))p(v|m(v))}\right),$$

is the mutual information (see Appendix B) between $\mathbf{v}$ and $\mathbf{k}$ given $\mathbf{m}$. To analyze further our privacy criteria the following definitions will be useful.

*Definition* 2.   Consider any three events $A$, $B$ and $C$. If $\Pr(A \wedge \bar{B}) \leq \gamma$, we say that the event $A$ probabilistically imply the event $B$ except with probability $\gamma$, and we write $A \Rightarrow_\gamma B$. If we have $\Pr(A \wedge \bar{B}|C) \leq \gamma$, we write $A \Rightarrow_{\gamma|C} B$.

One can easily check the following proposition:

PROPOSITION 1.   *For any events $A$, $B$ and $C$ we have (1) $A \Rightarrow_\gamma B$ and $B \Rightarrow_{\gamma'} C$ implies $A \Rightarrow_{\gamma+\gamma'} C$, (2) if $A \Rightarrow_\gamma B$ and $A \Rightarrow_{\gamma'} C$, then $A \Rightarrow_{\gamma+\gamma'} (B \wedge C)$, and (3) if $A \Rightarrow_\gamma B$, then $A \wedge C \Rightarrow_\gamma B$. Also $A \Rightarrow_{\gamma|C} B$ is equivalent to $A \wedge C \Rightarrow_{\gamma\times\Pr(C)} B$.*

*Definition* 3.   Consider any number $\sigma \geq 0$. Eve's view $v$ in a QKD protocol is $\sigma$-informative about $k$ if $|p(k|v) - 1/2^m| \leq 2^{-m} \sigma$. We denote by $\mathcal{N}_\sigma$ the event which is TRUE whenever the view $v$ is $\sigma$-informative about $k$.

The following lemma connects Definition 3 with our definition of privacy.

LEMMA 1.   *For every $\sigma > 0$, $\xi > 0$ and $m^{max} > 0$, if we have (1) $(\forall v)\ m(v) \leq m^{max}$ and (2) $\mathcal{P} \Rightarrow_\xi \mathcal{N}_\sigma$, then the protocol is f-private with $f = m^{max}\,\xi + \sigma/ln\ 2$.*

PROOF.   Let $\mathcal{I} = \mathcal{N}_\sigma \wedge \mathcal{P}$. Using $\mathcal{P} \Rightarrow_\xi \mathcal{N}_\sigma$ we obtain $\mathcal{P} \Rightarrow_\xi \mathcal{I}$. Using (2), we obtain

$$\sum_m p(m)(m - H_m(\mathbf{k}|\mathbf{v}))$$

$$= \sum_{k,v} p(k,v)(\mathbf{m}(v) + \log_2 p(k|v))$$

$$= \sum_{k,v|\mathcal{P}} p(k,v)(\mathbf{m}(v) + \log_2 p(k|v))$$

$$+ \sum_{k,v|\bar{\mathcal{P}}} p(k,v)(\mathbf{m}(v) + \log_2 p(k|v)).$$

The second term vanishes because $v \in \bar{\mathcal{P}}$ implies (1) $m(v) = 0$ and (2) $\log_2 p(k|v) = \log_2 p(k|v) = \log_2 1 = 0$ since there is only a single value for the null string. (We recall that we adopted the convention that the key $\mathbf{k}$ is set to the null

string when the test fails.) Therefore, we have

$$\sum_m p(m)(m - H_m(\mathbf{k}|\mathbf{v}))$$

$$= \sum_{k,v|\mathscr{P}} p(k, v)(m(v) + \log_2 p(k|v))$$

$$= \sum_{(k,v)|\mathscr{I}} p(k, v)(m(v) + \log_2 p(k|v))$$

$$+ \sum_{(k,v)|\mathscr{P}\wedge\bar{\mathscr{I}}} p(k, v)(m(v) + \log_2 p(k|v)).$$

We bound the first term of the last equation, the sum over $(k, v)$ such that $\mathscr{I}(k, v)$ is TRUE, via the relation

$$\mathscr{I}(k, v) \Rightarrow p(k|v) = 2^{-m}(1 + \sigma_{k,v}),$$

where $|\sigma_{k,v}| \leq \sigma$. For the second term, the sum over $(k, v)$ such that $\mathscr{P}(v) \wedge \mathscr{I}(k, v)$ is TRUE, we will drop the nonpositive terms $p(k, v) \log_2 p(k|v)$. We get

$$\sum_m p(m)(m - H_m(\mathbf{k}|\mathbf{v})) \leq \sum_{(k,v)|\mathscr{I}} p(k, v)\log_2(1 + \sigma_{k,v})$$

$$+ \sum_{(k,v)|\mathscr{P}\wedge\bar{\mathscr{I}}} p(k, v)m(v).$$

We finally obtain

$$\sum_m p(m)(m - H_m(\mathbf{k}|\mathbf{v})) \leq \frac{\sigma}{\ln 2} + m^{\max}\xi$$

where we used the inequality $\log_2(1 + x) \leq |x|/\ln 2$ for any $x > -1$ to obtain the first term, and $\Pr(\mathscr{P} \cap \mathscr{I}) \leq \xi$ and $m(v) \leq m^{\max}$ to obtain the second term. This concludes the proof. $\square$

## 3. *Some Useful Tools*

This section provides a general model and techniques that are useful to analyze the security of our quantum protocol.

3.1. A FICTIVE TEST LEMMA. Assume that Alice transmits some arbitrary binary string $g \in \{0, 1\}^N$ to Bob over a channel, quantum or classical. Let $D \subseteq \{1, \ldots, N\}$ be any subset of the positions. For example, if the channel is lossy, $D$ could be the set of every position $i$ at which Bob detects a bit $h[i]$. In general, $D$ is the set of positions $i$ where Bob's bit $h[i]$ and Alice's bit $g[i]$ are presumed identical except for a small probability of error. A typical situation in quantum protocols is that Alice and Bob need some kind of indication that the number of errors in some subset $E \subseteq D$ is small and yet they cannot execute a test on $E$ because the bits in $E$ must remain private. To address this problem, they pick two random subsets $T$ and $E$ so that the bits in $T$ can be used for a test while the bits in $E$ remain private. Every position $i \in D$ is added in $T$ (initially empty) with probability $p_T$ or in $E$ (initially empty) with probability $p_E$ or is ignored with probability $1 - p_T - p_E$. (In our protocol, we will deal with the case $p_T + p_E = 1$, but here we are slightly more general.)

The following lemma is a variation on the principle that says that the number of errors that one detects in $T$ is a good indication of the number of errors that exists in $E$. It is unlikely that the number of errors is small in $T$ and large in $E$ at the same time.

LEMMA 2. *Consider a set of positions $D$ and a string of errors $e$ on $D$.* (*In the above scenario, the string $e$ is $g[D] \oplus h[D]$.) Let $T$ and $E$ be two random subsets of $D$ such that every position $i \in D$ is in $T$ with probability $p_T$ or in $E$ with probability $P_E$ or discarded with probability $1 - p_T - p_E$. For $X \in \{T, E\}$, let $n_X$ and $d_X$ be the size of $X$ and the number of errors in $X$ (the weight of $e[X]$) respectively. Denote $\mathcal{P}_T$ the event that $d_T < \delta p_T n_D$ where $\delta$ is some fixed parameter that represents the tolerated error rate. Denote $\mathcal{P}_E$ the event that $d_E < (\delta + \beta) p_E n_D$ where $\beta > 0$ is any positive real number. We have that*

$$\mathcal{P}_T \Rightarrow_{\mu(\beta, n_D)} \mathcal{P}_E \tag{3}$$

*where*

$$\mu(\beta, n_D) = exp\left( \frac{-\beta^2 \min\{p_T^2, p_E^2\}}{2\delta + \beta} n_D + \frac{2\beta^2 p_E^2}{(2\delta + \beta)^2} \right).$$

The term $2\beta^2 p_E^2/(2\delta + \beta)^2$ is independent of $n_D$ and can be ignored when $n_D$ is large. With regard to the above scenario, the lemma only uses the fact that $E$ and $T$ are random and not distinguishable until after the transmission. It provides an upper bound $\mu(\beta, n_D)$ on $\Pr(\mathcal{P}_T \wedge \mathcal{P}_E)$ for every variable fixed except $E$ and $T$. The upper bound still hold if we average over the other variables, except that we will not average over $n_D$ because the upper bound $\mu(\beta, n_D)$ itself depends upon $n_D$. The following lemma is a variation on Chernoff's lemma, a standard tool to deal with large numbers (e.g., see Kearns [1989]). It is the basic tool used in the proof of the fictive test lemma.

LEMMA 3 (CHERNOFF). *Let $X_1, \ldots, X_n$ be $n$ independent Bernoulli variables and let $S = \Sigma_{i=1}^n X_i$. If $Pr(X_i = 1) = p$ for $1 \leq i \leq n$, then for all $0 \leq \Delta p \leq 1$, we have*

$$Pr(S \geq n(p + \Delta p)) \leq exp(-2n(\Delta p)^2), \tag{4}$$

$$Pr(S \leq n(p - \Delta p)) \leq exp(-2n(\Delta p)^2). \tag{5}$$

PROOF OF THE FICTIVE TEST LEMMA. The basic idea of the proof is the following. The number of errors $d_D$ in $D$ is either (1) larger or equal to $\lceil (\delta + \beta/2) n_D \rceil$ or (2) smaller or equal to $\lfloor (\delta + \beta/2) n_D \rfloor$. In the first case, the probability of $\mathcal{P}_T$ is small. In the second case, the probability of $\mathcal{P}_E$ is small. In both cases, the probability of $\mathcal{P}_T \wedge \mathcal{P}_E$ is small.

We first do the case where $d_D \geq \lceil (\delta + \beta/2) n_D \rceil$. The condition $\mathcal{P}_T$ hold only if $d_T$, the number of errors in $T$, is strictly smaller than $\delta p_T n_D$. We will bound from above the probability of $d_T \leq \delta p_T n_D$. Every $i \in D$, in particular every position $i$ with $e[i] \neq 0$, belongs to $T$ with probability $p_T$. That is, each of the $d_D \geq (\delta + \beta/2)n_D$ errors is put in $T$ with probability $p_T$. We may conservatively assume that $d_D = \lceil (\delta + \beta/2)n_D \rceil$ because a larger value for $d_D$ will only decrease the probability. Let $S = d_T$ be the number of errors in $T$. We want to obtain an upper bound on $\Pr(d_T \leq \delta p_T n_D)$ using inequality (5). We substitute $n$ and $p$ in (5) by $d_D$ and $p_T$, respectively. We will find a value for $\Delta p$ such that we have

$$\delta p_T n_D \leq d_D(p_T - \Delta p), \qquad (6)$$

and thus $d_T \leq \delta p_T n_D$ implies $d_T \leq d_D(p_T - \Delta p)$ and (5) can be used to bound $\Pr(d_T \leq \delta p_T n_D)$. Since $(\delta + \beta/2)n_D \leq d_D$, we obtain (6) if we have $(\delta + \beta/2)n_D = \delta p_T n_D/(p_T - \Delta p)$. The solution for $\Delta p$ is $\Delta p = \beta p_T/(2\delta + \beta)$. Applying Chernoff's lemma with these values, we obtain that $d_T$ is smaller than $\delta p_T n_D$ (and thus smaller than $d_D(p_T - \Delta p)$) with a probability smaller or equal to

$$\exp\left(\frac{-2\beta^2 p_T^2}{(2\delta + \beta)^2} d_D\right).$$

Now, we use the fact $(\delta + \beta/2)n_D \leq d_D$ and obtain

$$\Pr(\mathcal{P}_T) \leq \mu_T(\beta, n_D) \overset{def}{=} \exp\left(\frac{-\beta^2 p_T^2}{2\delta + \beta} n_D\right).$$

The second case, $d_D \leq \lfloor (\delta + \beta/2)n_\mathcal{D} \rfloor$, is similar to the first case, except that we use inequality (4) instead of inequality (5) and $p = p_E$ instead of $p = p_T$. The event $\mathcal{P}_E$ occurs when $d_E$, the number of errors in $E$, is larger than $(\delta + \beta)p_E n_D$. Even though $d_D$, the total number of errors in $D$, is smaller than $\lfloor (\delta + \beta/2)n_D \rfloor$, we may assume that $d_D = \lfloor (\delta + \beta/2)n_D \rfloor$ respectively. Now, we find $\Delta p$ such that we have $d_D(p_E + \Delta p) \leq (\delta + \beta)p_E n_D$. Since $d_D \leq (\delta + \beta/2)n_D$, we obtain the seeked inequality if we have $(\delta + \beta/2)n_\mathcal{D} = (\delta + \beta)p_E n_D/(p_E + \Delta p)$. As in the first case, this equation has the solution $\Delta p = \beta p_E/(2\delta + \beta)$. Using (4), we get $d_T$ is larger than $(\delta + \beta)p_T n_D$ (and thus larger than $d_D(p_E + \Delta p)$) with a probability smaller or equal to

$$\exp\left(\frac{-2\beta^2 p_E^2}{(2\delta + \beta)^2} d_D\right).$$

Here, unlike the first case, we don't have $(\delta + \beta/2)n_D \leq d_D$. We only have $(\delta + \beta/2)n_D - 1 \leq d_D$. We can do as in the first case, except for an additional positive term $2\beta^2 p_E^2/(2\delta + \beta)^2$ in the exponent. So we obtain

$$\Pr(\bar{\mathcal{P}}_E) \leq \mu_E(\beta, n_D) \overset{def}{=} \exp\left(\frac{-\beta^2 p_E^2}{2\delta + \beta} n_D + \frac{2\beta^2 p_E^2}{(2\delta + \beta)^2}\right).$$

We have

$$\Pr(\mathcal{P}_T \wedge \bar{\mathcal{P}}_E) \leq \mu(\beta, n_D) = \max\{\mu_T(\beta, n_D), \mu_E(\beta, n_D)\}.$$

This concludes the proof. □

3.2. ON PRIVACY AMPLIFICATION. A standard technique for privacy amplification is to extract a smaller key $k = K \cdot g$ of length $m$ out of a nonprivate string $g \in \{0, 1\}^n$ where $K$ is a $m \times n$ binary matrix and the matrix multiplication use the sum modulo 2. As explained in Bennett et al. [1988] one must pick the matrix $K$ adequately in view of how much information did leak out about $g$, including extra information that might have been announced for error correction. However, the results obtained in Bennett et al. [1988] do not apply if information can leak about $g$ in view of $K$ that, as we will see, is the case in our protocol, and perhaps in other quantum protocols. We use a different approach.

Let us consider a typical situation where extra information $s = F \cdot g$, where $F$ is a $r \times n$ binary matrix, is provided about $g$ for error correction (see Appendix C). Let $G^*$ be the set of linear combinations of rows in $F$ and $K$, which contains at the least one row in $K$, that is, $G^* = C^\perp[F, K] - C^\perp[F]$ where $C^\perp[F, K]$ is the set of linear combinations of rows in $F$ and $K$, and $C^\perp[F]$ is the set of linear combinations of rows in $F$ only (see Appendix C). We denote by $d_W$ the minimal weight of the strings in $G^*$. For example, if the parity check matrix $F$ is the $2 \times 5$ matrix with the two rows [10000] and [01000] and the privacy matrix $K$ is the $1 \times 5$ binary matrix with the single row [11111], we obtain that $G^* = \{[11111], [01111], [10111], [00111]\}$ so that $d_W = 3$. It is not hard to see that if Eve gets a bit $g[i]$ in less than $d_W$ positions $i$, and nothing about the bits at other positions, she learns nothing about the key $K \cdot g$ even after she receives the syndrome $s = F \cdot g$. The case where Eve learns bits at given positions and nothing at other positions, though it is not sufficient to understand the principle in its full generality, at the least suggests that privacy requires a large value for $d_W$. So, it is not surprising that in our protocol, which as we will see uses this kind of privacy amplification technique, we will need a lower bound on $d_W$. Note also that $d_W > 0$ implies that the rows of $K$ are linearly independent which is important for privacy. Usually, $d_W$ is much larger than 3. The following lemma provides one way to obtain a lower bound on $d_W$.

LEMMA 4. *Consider any $\tau > 0$ and any fixed $r \times n$ parity check matrix $F$. Let $K$ be a randomly chosen $m \times n$ binary matrices. The minimal weight $d_W$ of $C^\perp[F, K] - C^\perp[F]$ is greater than $H^{-1}(1 - (m + r)/n - \tau)n$ with a probability greater than $1 - \lambda$ where $\lambda = 2^{-\tau n}$.*

*Remark.* The lemma can also be written as True $\Rightarrow_\lambda d_W \geq H^{-1}(1 - (m + r)/n - \tau)n$

PROOF. Consider the sphere $\mathbf{S}$ of radius $d''' = H^{-1}(1 - [(r + m)/n] - \tau)n$ around the zero string $\mathbf{0}$. The minimal weight of $C^\perp[F] \oplus C^\perp[K]^*$ is smaller or equal to $d'''$ if and only if at the least one string in $C^\perp[F] \oplus C^\perp[K]^*$ belongs to $\mathbf{S}$. Therefore, it is sufficient to show that the probability $p$ that one string in $C^\perp[F] \oplus C^\perp[K]^*$ belongs to $\mathbf{S}$ is smaller than $2^{-\tau n}$. Let us analyze the random process by which $K$ is created. Let $w_i$ be the $i$th random row added to $K$. Let $K_i$ be the matrix which contains the rows $w_1, \ldots, w_i$ only. The probability that one string in $C^\perp[F] \oplus C^\perp[K]^*$ belongs to $\mathbf{S}$ is smaller than the sum (over $i$) of the probability $p_i$ that this happens when the string $w_i$ is added to $K$. Each probability $p_i$ takes its maximum value when all the strings $w_j$ with $j < i$ and the rows of $F$ are independent. When the random string $w_i$ is added, we have that one of $2^{n-r-i+1}$ disjoint sets $C^{\perp}[F] \oplus C^{\perp}[K_i]^*$ is being chosen perfectly at random. (These are the cosets of $C^{\perp}[F] \oplus C^{\perp}[K_{i-1}]$). The probability $p_i$ is larger when no two strings in $\mathbf{S}$ belongs to the same set $C^{\perp}[F] \oplus C^{\perp}[K_i]^*$. In this extreme case, the number of sets $C^{\perp}[F] \oplus C^{\perp}[K_i]^*$ which contains a string in $S$ reaches its maximal value, which of course is $|S|$. Using Lemma 8, the number of strings in $\mathbf{S}$ is smaller than $2^{H(d'''/n)n} = 2^{n-r-m-\tau n}$. Therefore, we have

$$p_i \leq \frac{2^{n-r-m-\tau n}}{2^{n-r-i+1}} = 2^{-m-\tau n + (i-1)}$$

Summing over $i = 1, \ldots, m$, we get

$$p \leq 2^{-m-\tau n} \sum_{i=0}^{m-1} 2^i \leq 2^{-\tau n}$$

This concludes the proof. $\square$

3.3. QUANTUM PRELIMINARIES. Quantum states are denoted by Greek letters $\psi$, $\phi$, etc. To lighten the notation, except in some occasions, we do not use the ket notation "$|\cdot\rangle$" around quantum states. A linear functional on $H$ is a linear transformation from $H$ to the space of complex numbers. The bra notation "$\langle\cdot|$" can be seen as the operation that sends a state $\phi$ into the unique linear functional $\langle\phi|$ such that, for every $\psi$, the value of $\langle\phi|$ evaluated at $\psi$ is the scalar product $\langle\phi|\psi\rangle$ between $\phi$ and $\psi$. We often denote this unique linear functional $\phi^{\dagger}$ rather than $\langle\phi|$. The tensor product of any two quantum states $\psi$ and $\phi$ is written as $\psi\phi$ or $\psi \otimes \phi$.

A quantum system is more than an abstract Hilbert space. It is an actual quantum system in the protocol. Two distinct quantum systems might have the same abstract Hilbert space. The ordering of the states in an expression such as $\psi_1 \otimes \psi_2 \in H_1 \otimes H_2$ is important, but this ordering can be seen as an association between the systems $H_1$ and $H_2$ and their respective states $\psi_1$ and $\psi_2$, and therefore not always related to the respective positions of $\psi_1$ and $\psi_2$ in the notation. In particular, if the association between the systems $H_i$ and their respective state $\psi_i$ is nonambiguous from the context, then $\psi_1 \otimes \psi_2$ and $\psi_2 \otimes \psi_1$ represent one and the same product state with the same state space association. If $E$ is an operator on a system $H_1$ and $\psi$ is a state that belongs to a product $H_1 \otimes H_2$, we adopt the convention that $E\psi = (E \otimes \mathbf{I}_{H_2})\psi$. Similarly, for every $\psi_1 \in H_1$ and $\phi = \psi'_1 \otimes \psi_2 \in H_1 \otimes H_2$, we adopt the convention that $\langle\psi_1|\phi\rangle = \langle\psi_1|\psi'_1\rangle|\psi_2\rangle \in H_2$, and this convention can be extended to nonproduct states $\phi \in H_1 \otimes H_2$ by linearity. In general, every operator $E$ on a state space $H$ can be written as a linear combination of basic operators $|\phi\rangle\langle\phi'|$ with $|\phi\rangle$, $|\phi'\rangle \in H$. Therefore, the previous rule can be used to define $\langle\psi|E|\psi'\rangle$ when $E$ is an operator on $H = H_1 \otimes H_2$ and $|\psi\rangle$, $|\psi'\rangle$ belong to $H_1$.

Consider the preparation of a state $|\psi_c\rangle$ with probability $p(c)$. As usual, if the random value $c$ is kept private, except for what can be learn via measurements on the state $|\psi_c\rangle$, this preparation is conveniently represented by the density operator $\rho = \Sigma_{c \in V} \, p(c)|\psi_c\rangle \, \langle\psi_c|$.

The partial trace over the system $H_1$, denoted $\text{Tr}_{H_1}$, is a linear mapping from the space of linear operators on $H_1 \otimes H_2$ to the space of linear operators on $H_2$ which maps any operator $E$ on $H_1 \otimes H_2$ into $\Sigma_i \langle \psi_i | E | \psi_i \rangle$ where $\{ | \psi_i \rangle \}_{i=1}^n$ is any orthoglobal basis of $H_1$. The (partial) trace operation over the entire system is denoted $\text{Tr}$. It is a linear functional on the space of operators. The identity $\text{Tr}_H(A | \psi \rangle \langle \psi |) = \langle \psi | A | \psi \rangle$, in which $A$ is any operator on $H$ and $| \psi \rangle$ any state in $H$, is often used.

Any physical process can be seen as a black box which receives an initial state $\psi$ of some quantum system $H$ and returns two components: a random classical outcome $x$ and an associated final quantum state $\psi_x$ that lies in some other quantum system $H'$. The POVM formalism [Peres 1993] describes this kind of process via a mapping $x \mapsto E_x$ where $x$ represents the random classical outcome of the process and $E_x$: $H \mapsto H$ is a positive operator on $H$ called the *measurement operator* associated with $x$. Note that the POVM formalism ignores the final system for the residual state, so $H'$ is not in the formalism. The mapping $x \mapsto E_x$ is only a formal trick to compute the probability of the classical outcomes $x$. If the initial density operator state is $\rho$, then the probability of $x$ is $\text{Tr}(E_x \rho)$. The final state $\rho_x$ is not given by the POVM model (and this is not a problem because we don't need to compute any residual state in the proof). Let $\mathbf{y}$ be a deterministic function of the outcome $x$. Note that $y$ can also be seen as the outcome of a measurement since the computation of $y$ is also a physical process. The measurement operator associated with $y$ is

$$E_y = \sum_{x | \mathbf{y}_{\mathbf{x}}(x) = y} E_x. \tag{7}$$

(For a definition of $\mathbf{y}_{|\mathbf{x}}$ see the subsection about random variables in Section 2.) We will extend a little bit the formalism to encompass the fact that a POVM on a system $H$ can be executed in view of some classical information $y \in Y$ which is available before the POVM is executed. For every $y$, we denote by $x \mapsto E_{x|y}$ or simply $E_{x|y}$ the POVM which is executed on a system $H$ in view of $y$ and returns an outcome $x$. We say that this conditioned POVM is executed on $H \times Y$. This notation is useful when the POVM is executed in view of some classical information $y$ available before the POVM is executed or when some previous POVM returned an outcome $y$ in view of which the POVM is executed. It is well known that, in accordance with the basic axioms of quantum mechanics, any physical process can be described with such a formalism [Peres 1993].

The conjugate transpose $E^\dagger$ of an operator $E$ is obtained by first transposing its matrix representation and then complex conjugating the entries. It can be shown, that this definition is independent of the basis that we use to represent the operator as a matrix. We have that $(E_1 E_2)^\dagger = E_2^\dagger E_1^\dagger$.

For every POVM $E_{x|y}$ on $H$, there exist a collapse operation $A_{x|y}$ on $H$ such that $E_{x|y} = A_{x|y}^\dagger A_{x|y}$. Suppose that some POVM $E_{y|c}^1$ is executed in view of some classical information $c$, and next another POVM $E_{x|y,c}^2$ is executed in view of $c$ and the outcome $y$ of the first POVM. Let $A_{y|c}^1$ and $A_{x|y,c}^2$ be collapse operations associated with the first and the second POVMs, respectively. The collapse operation associated with the overall POVM that describes the effect of both POVMs is $A_{x,y|c} = A_{x|y,c}^2 A_{y|c}^1$. Therefore, the overall POVM is

$$E_{x,y|c} = A_{x,y|c}^{\dagger} A_{x,y|c}$$
$$= A_{y|c}^{1\dagger} \mathcal{A}_{x|y,c}^{2\dagger} \mathcal{A}_{x|y,c}^{2} \mathcal{A}_{y|c}^{1}$$
$$= A_{y|c}^{1\dagger} E_{x|y,c}^{2\dagger} \mathcal{A}_{y|c}^{1}.$$

Note the strange fact that the overall POVM $E_{x,y|c}$ cannot be expressed directly in terms of the consecutive POVMs $E_{y|c}^{1}$ and $E_{x|y,c}^{2}$. This is one advantage of the collapse operation formalism over the POVM formalism.

3.4. THE BASIC MODEL. Here, we describe how to represent an attack against a quantum protocol in terms of the standard POVM formalism [Peres 1993]. The formalism should allow us to abstract the details and focus on the essential fact. In the standard model for quantum protocols, Alice and Bob use a set of registers and at every given step each register is controlled by one and only one participant. We assume that every transmitted register is first transmitted to the cheater. When a register is transmitted from a participant $X$ to a participant $Y$, the register that was controlled by $X$ is now controlled by $Y$. Since the cheater obtains control over the entire system that is transmitted, the details of the transmission can be safely ignored in the model.

We recall that we denote by $\hat{v}$ the overall random outcome of the protocol. This outcome fixes the value of every possible random variable in the protocol, including any classical announcement and any result of a quantum measurement. A basic principle in the model is that we will define a view $\mathbf{v}$ in the protocol (e.g., the view seen by a given participant) as a function $\mathbf{v} = \mathbf{v}(\hat{v})$ of this overall outcome. As we will see, our model separates the overall outcome $\hat{v}$ in two parts: a classical part $\hat{c}$ that corresponds to a random tape and a quantum part $\hat{q}$ which correspond to the outcome of the overall quantum measurement executed jointly by all participants in the protocol. So a view is a function $\mathbf{v}$ on the overall outcome $\hat{v} = (\hat{c}, \hat{q})$.

In a way, the separation of $\hat{v}$ in two components $\hat{c}$ and $\hat{q}$ is artificial because it is always possible to represent every bit in the random tape $\hat{c}$ as the outcome of a quantum measurement, a quantum cointoss. In this way, every register, classical or quantum, could be considered as if it was a quantum register. However, in some part of our analysis it is convenient to use methods of computation that are naturally understood in terms of classical information, and thus we don't want to always think in terms of quantum information. In any case, the protocol is described in terms of classical information as well as quantum information and at some point in the proof one must refer to the classical part of the protocol. Therefore, a careful understanding of the connection between the classical part and the quantum part is required. Such a connection is not as trivial as one might first think.

The state space for the quantum registers is denoted $H^{Q}$. The quantum system $H^{Q}$ contains every quantum register that could eventually be sent in the protocol as well as any register that could be measured or transformed jointly with such a register. Therefore, $H^{Q}$ is never entangled with another quantum system. We denote by $\hat{c}$ the content of classical random tapes or random registers available at the beginning of the protocol. We denote by $\hat{C}$ the set of possible values for $\hat{c}$. The random variable $\hat{\mathbf{c}}$ has some apriori distribution of probability $p(\hat{c})$. The only operation allowed on $\hat{C}$ in our context is the computation of some deterministic function $\mathbf{x}$ of $\hat{c}$, but there are no constraint on the function $\mathbf{x}$.

We denote by $\hat{q}$ the outcome of all measurements which are executed on $H^Q$ in view of $\hat{c}$. Without loss of generality we assume that all random values in the protocol can be computed deterministically from $\hat{v} \overset{def}{=} (\hat{c}, \hat{q})$. A view $\mathbf{v}$ in the protocol is any deterministic function of $\hat{v}$. A view is thus a random variable. We adopt the convention that $\mathbf{v} = v$ means $\mathbf{v}(\hat{v}) = v$. So, there is a global view $\hat{v} = (\hat{c}, \hat{q})$ defined by the protocol, and each participant's view is simply a deterministic function $\mathbf{v}$ on $\hat{v}$ which returns the specific information that belongs to this participant. This function can be deterministic because the random tape, and every thing else that is random in the view, is already included in the global view $\hat{v}$.

We denote by $E_{\hat{q}|\hat{c}}$ the conditioned POVM on $H^Q \times \hat{C}$ that is associated with $\hat{v} = (\hat{c}, \hat{q})$. We recall that the notation $E_{q|c}$ means that the POVM is executed on $H^Q$ in view of $c$ and returns an outcome $q$. Note that by definition of $H^Q$ such a POVM exists. For a given value $\hat{c}$, the overall initial state of the protocol is $|\Psi(\hat{c})\rangle^Q = \otimes_X |\Psi_X(\hat{c})\rangle^{Q_X}$, a pure state jointly prepared by the participants $X =$ Alice, Bob, . . . at the beginning of the protocol in view of their share of the random value $\hat{c}$. The probability of the initial state $|\Psi(\hat{c})\rangle^Q$ with $\hat{c} \in \hat{C}$, is $p_{\hat{c}}(\hat{c})$. In accordance with the POVM formalism, the probability of $\hat{v} = (\hat{c}, \hat{q})$ given this initial mixture is

$$p_{\hat{v}}(\hat{c}, \hat{q}) = p_{\hat{c}}(\hat{c}) \text{Tr}_Q(E_{\hat{q}|\hat{c}} |\Psi(\hat{c})\rangle\langle\Psi(\hat{c})|). \qquad (8)$$

3.4.1. *The Natural Separation Quantum Vs Classical*.  We want to provide formula that respect the natural separation between the classical part and the quantum part in a protocol. The idea is to take advantage of our classical intuition whenever possible when we compute probabilities. For this purpose, it will be convenient to consider views that can be written in the form $\mathbf{v} = (\mathbf{c}, \mathbf{q})$ where

C1 the deterministic function $\mathbf{c}$ depends on $\hat{c}$ and $q$ only, that is, it depends on $\hat{q}$ only via $q = \mathbf{q}(\hat{c}, \hat{q})$ and

C2 the POVM $E_{q|\hat{c}}$ on $H^Q$ defined via

$$E_{q|\hat{c}} \overset{def}{=} \sum_{\hat{q}|\mathbf{q}(\hat{c}, \hat{q}) = q} E_{\hat{q}|\hat{c}}$$

respects $E_{q|\hat{c}} = E_{q|c}$, that is, the mapping $(q, \hat{c}) \mapsto E_{q|\hat{c}}$ depends on $\hat{c}$ only via $c = \mathbf{c}(\hat{c}, q)$.

These two conditions allow us to take a point of view where in someway classical computations and quantum measurements are done separately. Condition C1 says that given the outcome $q$ of the POVM $E_{q|\hat{c}}$ one can compute the classical part $c$ as a function of the random tape $\hat{c}$ only (i.e., without having to use the state space $H^Q$). Condition C2 is similar, but it swaps the role of the random tape and the state space. These two conditions are respected in the following typical situation. First, ignoring $\hat{c}$, a part $q_1$ of $q$ is obtained via a measurement on $H^Q$, then in view of $q_1$ some function $\mathbf{c}_1$ of $\hat{c}$ is computed and the result $c_1$ is included in $c$, then in view of $c_1$ another part $q_2$ of $q$ is obtained, and so on. These conditions are not respected by every partial view $v = (c, q)$. For example, suppose that $H^Q$ is the state space for a single photon and the random tape $\hat{c}$ contains the basis used to measure this photon. The outcome of the measurement is $\hat{q}$. If only the outcome of the measurement is announced then we have $q = \hat{q}$ and $c = \varepsilon$, the empty string. In this example, one does not know what basis was used to return the outcome $q = \hat{q}$ so the condition C2 is not respected. If later the basis $\hat{c}$ is announced, we obtain a new view where the basis $\hat{c}$ is part of the classical part, and condition C2 is respected. In our analysis, we will only consider views in which C1 and C2 are respected.

*Definition* 4. Let $\mathbf{v} = (\mathbf{c}, \mathbf{q})$ be any view that respects C1. We define

$$p(c:q) \stackrel{def}{=} \sum_{\hat{c}|\mathbf{c}(\hat{c},q) = c} p(\hat{c}).$$

Note that $p(c:q)$ is not identical to $p(c|q)$. The following proposition is very easy to prove, but is nevertheless useful.

PROPOSITION 2. *Let* $\mathbf{v} = (\mathbf{c}, \mathbf{q})$ *be any view on* $\hat{v}$ *that respects conditions C1 and C2. We have*

$$p(v) = p(c:q)Tr_Q(E_{q|c}\rho_{c|q}), \tag{9}$$

*where*

$$\rho_{c|q} \stackrel{def}{=} p(c:q)^{-1} \sum_{\hat{c}|\mathbf{c}(\hat{c},q) = c} p(\hat{c})|\Psi(\hat{c})\rangle\langle\Psi(\hat{c})| \tag{10}$$

*is the renormalized density matrix on* $H^Q$ *associated with the constraint* $\mathbf{c}(\hat{c}, q) = c$ *on* $\hat{C}$.

PROOF. The probability of the view $v = (c, q)$ is given by

$$p_{\mathbf{v}}(c, q) = \sum_{\hat{c}|\mathbf{c}(\hat{c}, q) = c} \sum_{\hat{q}|\mathbf{q}(\hat{c}, \hat{q}) = q} p_{\hat{c}}(\hat{c})Tr_Q(E_{\hat{q}|\hat{c}}|\Psi(\hat{c})\rangle\langle\Psi(\hat{c})|). \tag{11}$$

Note that without condition C1 the sum over $\hat{c}$ and the sum over $\hat{q}$ in Eq. (11) could not be separated in this way. Now, the trace operation can be taken in evidence in front of these two sums. We obtain

$$p_{\mathbf{v}}(c, q) = Tr_Q\left(\sum_{\hat{c}|\mathbf{c}(\hat{c},q) = c} p_{\hat{c}}(\hat{c}) \sum_{\hat{q}|\mathbf{q}(\hat{c},\hat{q})=q} E_{\hat{q}|\hat{c}}\left|\Psi(\hat{c})\right\rangle\left\langle\Psi(\hat{c})\right|\right).$$

Because of C2, the sum $\Sigma_{\hat{q}|\mathbf{q}(\hat{c},\hat{q})=q} E_{\hat{q}|\hat{c}}$ is the operator $E_{q|c} = E_{q|c}$. The operator $E_{q|c}$ can be taken in evidence out of the sum over $\hat{c}$. We obtain

$$p_{\mathbf{v}}(c, q) = Tr_Q\left(E_{q|c} \sum_{\hat{c}|\mathbf{c}(\hat{c},q) = c} p_{\hat{c}}(\hat{c})\left|\Psi(\hat{c})\right\rangle\left\langle\Psi(\hat{c})\right|\right).$$

After renormalization of the density matrix one obtains the result. This concludes the proof. □

*Remark.* We emphasize the importance of getting the intuition of what is going on when we compute $\rho_{c|q}$ and $p(c:q)$ which are needed in Proposition 2.

Let $\mathbf{c}_q$ be the function on $\hat{c}$ defined via $\mathbf{c}_q(\hat{c}) = \mathbf{c}(\hat{c}, q)$. The density matrix $\rho_{c|q}$ is exactly the density matrix that you obtain when you prepare $\Psi(\hat{c})$ with the apriori probability $p_{\hat{c}}(\hat{c})$ then compute $\mathbf{c}_q$ and only keep the states for which $\mathbf{c}_q = c$. This fact is useful to write down an explicit expression for $\rho_{c|q}$. Now, we consider the probability $p(c : q)$, that is, the probability that $\mathbf{c}_q = c$. Since we defined $\mathbf{c}_q$ as a function on $\hat{c}$, a formal and complicated way to compute the probability $p(c : q)$ would be to determine $\mathbf{c}_q^{-1}(c) = \{\hat{c} | \mathbf{c}_q(\hat{c}) = c\}$, the pre-image of $c$, and then compute the probability of this set using the a priori probability of $\hat{c}$. However, given $q$ fixed, $\mathbf{c}_q$ is a random variable that is typically defined by the protocol in terms of only a few components of $\hat{c}$ and it will be simple to compute $p(c : q)$.

3.4.2. *Extended Operator Formalism.* In a few situations it turns out to be useful to consider classical registers as a special case of quantum registers. We denote by $|\hat{c}\rangle^C$, $\hat{c} \in \hat{C}$, the associated orthogonal states. The state space for the classical part is denoted $H^C$. The initial random state is $|\hat{c}\rangle^C |\Psi(\hat{c})\rangle$ with probability $p(\hat{c})$ and we denote by $\rho$ the associated density matrix. The POVM that returns $(\hat{c}, \hat{q})$ is $E_{\hat{v}} = P_{\hat{c}} E_{\hat{q}|\hat{c}}$ where $P_{\hat{c}}$ is the projection on the state $|\hat{c}\rangle^C$. In particular, we have

$$p(\hat{v}) = \mathrm{Tr}(E_{\hat{v}}\rho) = p(\hat{c})\langle \Psi(\hat{c}) | E_{\hat{q}|\hat{c}} | \Psi(\hat{c})\rangle,$$

which is consistent with (8). In our proof, we will only need the extended operator formalism to use the simple rule $E_z = \Sigma_{v | \mathbf{z}(v) = z} E_v$ which is valid for any deterministic function $\mathbf{z}$ of $v = (\mathbf{c}, \mathbf{q})$. We will use this rule in a context where $E_v$ is not used to compute a probability. In such a context, at the least from the author point of view, it is hard to explain this rule without the extended operator formalism.

However, we don't want to use the extended operator formalism at every step. We now explain how to pass from the standard to the extended operator formalism. The following proposition is easily obtained:

PROPOSITION 3. *Let $\mathbf{v} = (\mathbf{c}, \mathbf{q})$ be a view on $\hat{v}$ that respects C1 and C2. We have that the POVM on $H^C \otimes H^Q$ associated with $v = (c, q)$ is $E_v = E_{(c,q)} = P_{c|q}E_{q|c}$ where*

$$P_{c|q} = \sum_{\hat{c} | \mathbf{c}(\hat{c}, q) = c} P_{\hat{c}}.$$

PROOF. Using condition C1, we obtain

$$E_v = E_{(c,q)} = \sum_{\hat{c} | \mathbf{c}(\hat{c}, q) = c} \; \sum_{\hat{q} | \mathbf{q}(\hat{c}, \hat{q}) = q} P_{\hat{c}} E_{\hat{q}|\hat{c}}$$

$$= \sum_{\hat{c} | \mathbf{c}(\hat{c}, q) = c} P_{\hat{c}} \sum_{\hat{q} | \mathbf{q}(\hat{c}, \hat{q}) = q} E_{\hat{q}|\hat{c}}$$

Using condition C2, we obtain

$$E_v = \sum_{\hat{c} | \mathbf{c}(\hat{c}, q) = c} P_{\hat{c}} E_{q|c} = P_{c|q} E_{q|c}.$$

This concludes the proof. □

Let $\Pi$ and $\Pi'$ be any two operators on $H^Q$. We need these operators to make our next equation sufficiently general for our purpose. The following equation will be useful in our proof to pass from one formalism to the other.

$$\mathrm{Tr}(E_v\Pi\rho\Pi') = \mathrm{Tr}(P_{c|q}E_{q|c}\Pi\rho\Pi')$$

$$= p(c\!:\!q)\mathrm{Tr}_Q(E_{q|c}\Pi\rho_{c|q}\Pi'). \tag{12}$$

The main content of this rule is in the second equality, the first equality being a direct consequence of Proposition 3. In the particular case $\Pi = \Pi' = \mathbf{I}_E$, since $\mathrm{Tr}(E_v\rho) = p(v)$, this rule is essentially Proposition 2. In Proposition 2, we have that the sum over $\hat{c}$ in the definition of $\rho_{c|q}$ is restricted by the condition $\mathbf{c}(\hat{c}, q) = c$. The basic point of formula (12) is that in the extended operator formalism this restriction is implemented via the projection $P_{c|q}$ on $\rho$, the factor $p(c\!:\!q)$ being added (in the nonextended formalism) to compensate for the fact that $\rho_{c|q}$ is normalized.

PROOF OF FORMULA (12). Note that $\mathrm{Tr} = \mathrm{Tr}_Q\mathrm{Tr}_C$, that is, the trace operation $\mathrm{Tr}$ corresponds to a partial trace over $H^C$ followed by a trace over $H^Q$. The operators $E_{q|c}$, $\Pi$ and $\Pi'$ commute with $\mathrm{Tr}_C$ and $P_{c|q}$ because the former operator on $H^Q$ whereas the latter operate on $H^C$. Therefore, acting on the left-hand-side, we can first execute $P_{c|q}$ and $\mathrm{Tr}_C$. If we expand $\rho$ as a sum over $\hat{c}$, we obtain that the restriction on $\hat{c}$ that comes with the projection $P_{c|q}$ (the same restriction as in Proposition 2) followed by the partial trace $\mathrm{Tr}_C$ maps $\rho$ into $p(c\!:\!q)\rho_{c|q}$. The factor $p(c\!:\!q)$ is needed to compensate for the fact that $\rho_{c|q}$ is renormalized. After taking $p(c\!:\!q)$ in evidence, we are left with the right-hand side. □

## 4. *The Protocol*

To focus on the basic procedure, we first describe the protocol without the validation constraints. Then we describe the validation constraints on the length of the key, etc. Next, the maximum error rate that we can tolerate in the protocol and yet obtain a key of non-zero length is derived from the validation constraints.

4.1. THE PROTOCOL. The protocol analyzed is a variation on the well known protocol proposed by Bennett and Brassard [1984] (see also Bennett et al. [1992]).

*Step* 1. *Alice's Preparation*. Alice sends $N$ two dimensional quantum systems to Bob prepared individually in one of the four BB84 states uniformly picked at random. For concreteness, one can think that the BB84 states denoted $\Psi(0, +)$, $\Psi(1, +)$, $\Psi(0, \times)$ and $\Psi(1, \times)$ correspond to photons polarized at 0, 90, 45, and $-45$ degrees, respectively. Of course, the proof remains the same if Alice uses any other realization of the BB84 states described in Figure 1. The state of the photons prepared by Alice is $\Psi(g, a) \overset{def}{=} \Psi(g[1], a[1]) \otimes \cdots \otimes \Psi(g[N], a[N])$.

For any string of bits $\alpha \in \{0, 1\}^E$ and string of bases $\theta \in \{+, \times\}^E$ (see Appendix C), the state $|\Psi(\alpha, \theta)\rangle$ is a state for the photons in $E$ that encodes the bits $\alpha[i]$ in the bases $\theta[i]$ for each $i \in E$.

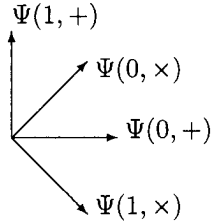

F<small>IG</small>. 1.  The BB84 states.

*Step* 2. *Bob's Measurement*.  Bob measures each photon using either the rectilinear basis {$\Psi(0, +)$, $\Psi(1, +)$} or the diagonal basis {$\Psi(0, \times)$, $\Psi(1, \times)$} uniformly chosen at random. If Bob detects a photon at position $i$, the associated outcome is denoted $\perp$ and we say that $i$ is a detected position. We adopt the following notation:

—$a \in \{+, \times\}^N$: Alice's string of bases.

—$g \in \{0, 1\}^N$: Alice's string of bits.

—$b \in \{+, \times\}^N$: Bob's string of bases.

—$\mathscr{D}$: the set of detected positions, that is, the set of positions $i$ with $h[i] \neq \perp$.

—$h \in \{0, 1, \perp\}^N$: Bob's string of outcomes.

*Step* 3. *Choosing the Tested Bits*.  Bob picks at random a subset of positions $R \subseteq \{1, \ldots, N\}$: he puts every position $i$ in the set $R$ (initially empty) with probability $p_T$.

*Some notations*.  We denote by $\Omega \subset \mathscr{D}$ the set of positions $i$ where $a[i] = b[i]$. Normally, except for errors in the transmission, for every $i \in \Omega$, we should have $g[i] = h[i]$. We denote by $T = R \cap \Omega = \{i \in R | a[i] = b[i]\}$ the set of tested positions.

*Step* 4. *Counting the Errors*.  Bob announces $R$ and $b$, Alice announces $a$ and $g[R]$, Bob announces $h[R]$. Alice and Bob note the value $d_T = d_T(g, h)$, the Hamming distance between $g$ and $h$ on $T$.

*Remark*.  The value $d_T$ will be used in validation constraints. These validation constraints specify valid values for the number of redundant bits needed for error-correction, the length of the key, etc.

*Step* 5. *Error Correction*.  The positions in $R$ are discarded. The set $E = \Omega - R = \{i \notin R | a[i] = b[i]\}$ will be used to define the key. Let $n_E = |E|$. If no error occurred, Alice and Bob should share the string $g[E]$ which we call the raw key. For error correction, Alice computes and announces to Bob the *syndrome* $s = F \bullet g[E]$ where $F$ is a $r \times n_E$ parity check matrix for a linear error-correcting code [MacWilliams and Sloane 1977], and "$\bullet$" is the binary matrix multiplication modulo 2 (see also Appendix C). The syndrome $s$ contains $r$ redundant bits about the raw key $g[E]$. The value $r$ depends on $d_T$ (see the validation constraints). Bob uses this information to correct the errors in $h[E]$ and obtain the raw key $g[E]$. More about error correction is provided in the next subsection.

*Step* 6. *Key Extraction.* At this point, Alice and Bob share the string $g[E]$ which we call the raw key. After error correction, to define a (final) key, Alice uses a $m \times n_E$ binary matrix $K$. The value of $m$ will depend on $d_T$. Alice and Bob compute the key $\hat{k} = K \bullet g[E]$. At this stage, in practice the protocol ends.

Eve is interested about the key $\hat{k} = K \bullet g[E]$ that is a function of the string $g$ chosen by Alice. In another variation on this protocol [Bennett et al. 1992]. Alice and Bob execute an interactive reconciliation procedure in which the parity check matrix $F$ is a function of the error positions. Such an approach where the matrix $F$ depends upon the error positions can certainly reduce the number of redundant bits needed in practice, but it makes the privacy proof more complicated.

4.1.1. *More about Error-Correction.* The normal way to use an error-correcting code with a $r \times n_E$ parity check matrix $F$ is to encode a message $x \in \{0, 1\}^k$ with $k = n_E - r$ into a codeword $c_x$ where by definition a codeword is a string $c$ such that $F \bullet c = 0$ (see Appendix C.2). The announcement of the syndrome $F \bullet g[E] = s$ by Alice for error-correction is not the normal way to use an error-correcting code with parity check matrix $F$ but it is convenient in the proof. In practice, Alice and Bob can use an alternative approach, which is closer to the normal way an error-correcting code is used, but is equivalent to the announcement of the syndrome $s$ from an information theoretic point of view. Let $\bar{g} = g[E]$ be the raw key. Before error-correction, Bob's received raw key is $\bar{h} = h[E]$. In this alternative approach, Alice chooses a random string $\mathbf{x} \in \{0, 1\}^k$ and compute the associated codeword $\mathbf{c}_x$ via the standard encoding procedure. She sends to Bob the redundant information $\hat{\mathbf{s}} = \bar{g} \oplus \mathbf{c}_x$. Note that a straightforward computation of the syndrome $s = F \bullet \bar{g}$ via matrix multiplication takes a number of steps in $O(n_E^2)$ and usually the encoding procedure is more efficient. Furthermore, in this alternative protocol, the error-correction procedure associated with the linear code can be used by Bob (together with two additions of the redundant information $\hat{\mathbf{s}}$) to obtain the string $\bar{g}$. Bob simply computes $\hat{\mathbf{s}} \oplus \bar{h}$ which is the codeword $c_x$ modulo the errors in $\bar{h}$. He does error-correction on this string to obtain $c_x = \hat{\mathbf{s}} \oplus \bar{g}$ and finally computes $\bar{g} = c_x \oplus \hat{\mathbf{s}}$.

In some cases, the encoding procedure adds the string $\bar{F} \bullet x$ to the left of a meaningful message $x \in \{0, 1\}^{n_E - r}$. In this case, one can check that $F = [\bar{F}\mathbf{I}_r]$ is a $r \times n_E$ parity check matrix for the linear code. Let $g[\text{mess}]$ be the first $n_E - m$ bits in $g$, and $g[\text{pari}]$ be the last $m$ bits. The syndrome associated with $F$ is $s = \bar{F} \bullet g[\text{mess}] \oplus g[\text{pari}]$. In this case, Alice can actually send the syndrome $s$ and Bob can compute $w = h[E] \oplus \hat{s}$ with $\hat{s} = [\mathbf{0}s]$, use the standard error-correction procedure on $w$ and add again $\hat{s}$ after error correction.

4.2. THE VALIDATION CONSTRAINTS. To complete the definition of the protocol, one must specify the validation constraints. In practice, it does not matter when these constraints are verified as long as they are verified before the key is used. However, in the proof, it will be assumed that the constraint are verified at the very end so that all variables are defined even if the constraints fail.

The basic validation constraint is $d_T < \delta p_T n_\Omega$ where $d_T = d_T(g, h)$ is the number of errors in $T$, $\delta > 0$ is some fixed parameter in the protocol and $n_\Omega = |\Omega|$. We denote by $\mathscr{P}_T$ the event that this constraint is satisfied. The tolerated number of errors is a constant fraction $p_T \delta$ of $n_\Omega$ rather than a constant fraction $\delta$ of $n_T \overset{def}{=} |T|$ because this particular choice was convenient in the proof of Lemma 2. Let $p_E = 1 - p_T$. Let us define

$$(\forall y \geq 0) \qquad d_+(y) \overset{def}{=} \lceil (\delta + y)p_E n_\Omega \rceil. \tag{13}$$

Lemma 2 with $D = \Omega$ applies directly to our protocol. Therefore, in a way, that is, in the sense given by Lemma 2, $d_+(\beta)$ is an upper bound on the number of errors that can be obtained on $E$ where $\beta$ is a parameter in the protocol with valid range $]0, \infty]$. We assume that, for each value of $n_E = |E|$, Alice and Bob know the number of errors $d'$ which can be corrected by their error-correction procedure. The validation constraint $\mathscr{X}_1$ for error-correction is $d' \geq d_+(\beta)$. Under the assumption that the error-correction procedure corrects up to $d'$ errors on $E$, it is not too difficult to conclude with the help of Lemma 2 that, except with an exponentially small probability (i.e., smaller than $\mu(\beta, n_\Omega)$), the cheater cannot succeed to have Alice and Bob think that they share a key when they do not.

Now, we give additional validation constraints for privacy. We recall that an important quantity is the minimal weight $d_W$ defined in Section 3.2. Privacy requires a large value for $d_W$. In our proof, we will see that the constraint $\mathscr{X}_2 : d_W \geq 2d_+(\epsilon)$, where $\epsilon$ is a parameter in the protocol with valid rangle $]0, \infty]$, is sufficient for privacy. However, one needs to compute $d_W$ to check this constraint. The matrices $F$ and $K$ are determined in view of $n_E$, $r$ and $m$, and therefore $d_W$ is a function of these parameters. If $d_W$ can be computed efficiently, we can use the validation constraint $\mathscr{X}_2$.

However, the evaluation of $d_W$ is in some cases a NP-complete problem [Barg 1997]. Let $H(x) = -[x \log_2(x) + (1 - x)\log_2(1 - x)]$ be the entropy function. If we cannot find a choice for $K = K_{m,n_E}$ and $F = F_{r,n_E}$ such that $d_W$ can be computed efficiently, we have an alternative approach to guarantee a lower bound on $d_W$. We will choose the $m \times n_E$ privacy matrix $K$ uniformly at random and adopt the constraint $\mathscr{X}_2'$ which says

$$H^{-1}\left(1 - \frac{r + m}{n_E} - \tau\right)n_E \geq 2d_+(\epsilon),$$

for some parameter $\tau > 0$. The parameter $\epsilon$ was defined before. As for $\epsilon$, the parameter $\tau$ can take any positive fixed value in the protocol. We recall that, given that the $m \times n_E$ privacy matrix $K$ is chosen uniformly at random, Lemma 4 tells us that, for any $\tau > 0$, we have

$$\text{True} \Rightarrow_\lambda d_W \geq H^{-1}\left(1 - \frac{r + m}{n_E} - \tau\right)n_E,$$

where $\lambda = 2^{-\tau n_E}$. Using Proposition 1 and what we have above, one easily obtains $\mathscr{X}_2' \Rightarrow_\lambda \mathscr{X}_2$. Therefore, as we will see, the criteria $\mathscr{X}_2'$ can replace the criteria $\mathscr{X}_2$. (The value $\lambda$ will contribute to the parameter $\xi$ that is used in Lemma 1.)

4.2.1. *The Validation Constraints*: *A Summary*.  The overall validation constraint is $\mathscr{P} \overset{def}{=} \mathscr{P}_T \wedge \mathscr{X}_1 \wedge \mathscr{X}_2' \wedge \mathscr{X}_3$ where $\mathscr{P}_T$: $d_T < \delta p_T n_\Omega$ where $\delta$ is the tolerated error rate (slightly above the expected error rate) and $p_T$ is the probability that any position $i \in \Omega$ is tested,

$\mathscr{X}_1$: $d' \geq d_+(\beta)$ where $p_E = 1 - p_T$, $d'$ is the number of errors which can be corrected by the error correction procedure and $\beta > 0$ is any positive value fixed in the protocol,

$\mathscr{X}'_2$: $H^{-1}(1 - (r + m)/n_E - \tau)n_E \geq 2d_+(\epsilon)$ where $\epsilon > 0$ and $\tau > 0$ are any positive values fixed in the protocol,

$\mathscr{X}_3$: $n_E \geq n_E^{\min}$, $n_\Omega \geq n_\Omega^{\min}$ and $m \leq m^{\max}$ where each of $n_E^{\min}$, $n_\Omega^{\min}$ and $m^{\max}$ are deterministic functions of all the parameters, and linear in the security parameter $N$.

Note that our privacy proof still hold even if the lower bounds $n_E^{\min}$ and $n_\Omega^{\min}$ are so big that they are unlikely to be respected. In fact, large lower bounds will just increase privacy. To avoid that the protocol fails most of the time, one must pick reasonably small lower bounds in view of $N$ and other parameters, but not for privacy. It is clear that reasonable lower bounds can be linear in $N$. A natural choice for $m^{\max}$ is the solution of $\mathscr{X}'_2$ for $m$, in which we use $\tau = -\Delta$, for some $\Delta > 0$ (say $\Delta = 1/10$), and replace $d_+(\epsilon)$, $n_E$ and $r$ by $d_+^{\mathrm{fair}}(\epsilon, p_E)$, $n_E^{\mathrm{fair}}$ and $r^{\mathrm{fair}}$, some fair estimate of their value chosen in view of $N$ and other parameters, respectively. It is clear that these estimates can be linear in $N$. The solution is

$$m^{\max} = n_E^{\mathrm{fair}} - r^{\mathrm{fair}} - H\left(\frac{2d_+^{\mathrm{fair}}(\epsilon, p_E)}{n_E^{\mathrm{fair}}}\right)n_E^{\mathrm{fair}} + \Delta n_E^{\mathrm{fair}}.$$

4.3. THE MAXIMUM TOLERATED ERROR RATE. The maximum tolerated error rate can be obtained as a function of the validation constraints. Only $\mathscr{X}'_2$ and $\mathscr{X}_1$ need to be considered because the other constraints do not restrict $\delta$ from above. More precisely, we want to find for which values of $\delta > 0$ these constraints have a significant probability to be satisfied as $N$ increases. Let us first consider $\mathscr{X}'_2$. For $N$ large, one can pick small values for $\epsilon$ and $\tau$. To compute the maximal value for $\delta$, we set $\epsilon = 0$ and $\tau = 0$ with the understanding that if we are actually below the maximal value for $\delta$, there will be room for positive value for these other parameters. So, we get $d_+(\epsilon) = \delta p_E n_\Omega$. For every $\epsilon' > 0$, in the limit of large $N$, we have that $|p_E n_\Omega/n_E - 1| \leq \epsilon'$ occurs with large probability. So we set $\epsilon' = 0$ or equivalently $p_E n_\Omega/n_E = 1$ with the same understanding as in the case of $\epsilon$ and $\tau$. So, we get $d_+(\epsilon) = \delta n_E$. Dividing by $n_E$ and applying $H$ on both sides of $\mathscr{X}'_2$, we obtain

$$1 - \frac{r + m}{n_E} \geq H(2\delta)$$

and thus

$$\frac{m}{n_E} \leq 1 - H(2\delta) - \frac{r}{n_E}.$$

Now, we must consider $\mathscr{X}_1$. Using the same principle as before, we set $\beta = 0$ and use $\epsilon' = 0$ to obtain that $d_+(\beta) = \delta p_E n_\Omega = \delta n_E$. Shannon's bound for error correction says that $r/n_E > H(\delta)$ is sufficient to correct $d'$ errors with $d' \geq \delta n_E$. So, if we use an error-correcting code that reaches Shannon's bound and $1 - H(2\delta) - H(\delta) \geq 0$, we have that, for sufficiently small value of $\epsilon$, $\beta$ and $\tau$, we can get $m/n_E \geq 0$ and satisfy $\mathscr{X}_1$ and $\mathscr{X}'_2$ with large probability in the limit of large $N$. Solving for $\delta$, we obtain that any value $\delta$ below 7.4% will do.

## 5. The Privacy Proof

Privacy in the protocol is expressed by the following theorem.

THEOREM 1. *Let $\delta > 0$ be the tolerated error rate and $p_T > 0$ be the probability that any given position $i \in \Omega$ is tested, that is, $\delta$ and $p_T$ are the parameters used in the validation constraint $\mathscr{P}_T$. Let $p_E = 1 - p_T$. Let $\epsilon > 0$ and $\tau > 0$ be the fixed parameters used in the validation constraint $\mathscr{X}_2$. Let $n_E^{min}$, $n_\Omega^{min}$ be the lower bounds and $m^{max}$ be the upper bound used in the validation constraint $\mathscr{X}_3$. Let $\mu$ be the following function of these parameters*

$$\mu = exp\left(\frac{-\epsilon^2 \, min\{p_T^2, p_E^2\}}{2\delta + \epsilon} n_\Omega^{min} + \frac{2\epsilon^2 p_E^2}{(2\delta + \epsilon)^2}\right).$$

*The same function that was defined in Lemma 2 except that here $\beta$ and $n_D$ are replaced by $\epsilon$ and $n_\Omega$ respectively. Let $\gamma = \mu^{1/2}$, $\eta = 2\sqrt{\gamma} + \gamma$, $\lambda = 2^{-\tau n_E^{min}}$, $\xi = \gamma + \lambda + \eta + 2\sqrt{2\eta}$ and $\sigma = \eta + \sqrt{2\eta}$. The protocol if f-private where $f = \sigma/ln(2) + m^{max}\, \xi$.*

This privacy result provides a bound on the amount of information that Eve can obtain about the final key. This bound holds as long as the length of the key is set by Alice and Bob in accordance with the validation constraints. As we mentioned earlier, the maximum value for the tolerated error rate $\delta$ is also determined by these validation constraints (see previous section).

On top of a perfect source, the only additional assumption required in the proof is, for every state of Bob's system $H^B$, the distribution of probability of $(\mathscr{D}, h[R])$ returned by Bob's measurement is the same whether this measurement uses the bases $b$ or the bases $\bar{b}$. We believe that this assumption is very reasonable. If we had that the measurements executed by Bob at different positions are independent (which is not too hard to obtain from an experimental point of view), we would only need the assumption that whether or not a photon is detected does not depend on the basis that is used to measure this photon. One can check that this alternative assumption together with the independence of Bob's measurements implies our assumption. Though it is not sufficient alone for our proof, this alternative assumption is the essential idea behind our assumption. This alternative assumption is always true, and thus not an assumption anymore, if no loss is tolerated in the transmission.

5.1. AN OVERVIEW OF THE PROOF AND SOME INTUITION. The main ingredient in the proof of privacy is that of complementarity. In our protocol, Alice encodes a string of bits in a certain choice of bases. From Alice's point of view, these bases are known and fixed at the beginning of the protocol. We call these bases the original bases. Eve does not know which bases are used by Alice and, as far as Eve is concerned, Alice could have used the opposite bases. The principle of complementarity tells that, if a measurement would give a lot of information about Alice's string had Alice chosen the opposite bases, the conjugate bases, then the same measurement can only provide little information about this string (which is encoded in the original bases).

Consider now the measurement that describes collectively the measurement executed by Eve and Bob on Alice's photons. The previous point implies that if Bob gets a lot of information about Alice's string had Alice chosen the conjugate bases, then Eve can only get little information about Alice's string encoded in the original bases. This suggests that we consider a scenario in which Alice uses the conjugate bases. However, that is not sufficient. To apply the complementary principle to this scenario, Bob must obtain a lot of information (when Alice uses the conjugate bases). Therefore, in this scenario, Bob must use the same bases as Alice, the conjugate bases. Considering such a scenario will allow us to use the complementary principle to show the security of a protocol that is slightly different from the original protocol, a protocol in which Bob uses the wrong bases. (Alice's bases return back to the original bases in the conclusion of the complementary principle.) This explains why an important aspect of our proof is the analysis of a *modified protocol* in which Bob uses the wrong bases. It also explains why the analysis of this modified protocol requires that we consider a scenario in which Alice also uses the wrong bases.

We said that to apply the complementary principle we must have that Bob obtains a lot of information. It is not sufficient that Bob uses the same bases as Alice to guarantee that he obtains a lot of information. For example, Eve could keep Alice's photons and send other photons to Bob. It is here that the test on a random set $R$ and the randomness of Alice's bases are important. As explained in Lemma 2, it is very unlikely in the above scenario (where Alice and Bob use the same bases) that the test on $R$ is successful and yet Bob obtains little information about Alice's bits outside $R$. So essentially there are two cases to consider: (1) the test fails and (2) Bob has a lot of information. We do not have to worry about what happens when the test fails. When the test is successful we apply the complementary principle.

Thus far, we explained how the complementary principle can be used to prove privacy in the modified protocol. To conclude, we need to return to the original protocol. Fortunately, as we will see, Bob's measurement on the nontested positions does not influence Eve's measurement and the classical data that is received by Eve, nor Alice's private key. Therefore, one can swap Bob's bases on the nontested positions back to their original orientation in the above conclusion without altering Alice's private key, nor Eve's knowledge on Alice's private key. However, we cannot swap Bob's bases on the tested positions without modifying Eve's view and, in fact, even the test result would not be identical. To solve this problem we will use the complementary principle as we explained above but only on the nontested positions, that is, only the bases in the nontested positions will be flipped in the modified protocol. Bob's modified string of bases (flipped outside $R$) will be denoted $\bar{b}$. As we will see, this partial flip of bases will not be a problem.

Formally, the proof will look as follows: First, privacy in the protocol is reduced to privacy in the modified protocol where Bob uses the flipped bases $\bar{b}$. Second, we apply our general model to the modified protocol to obtain a POVM description of the attack. Third, privacy in the modified protocol is proven by considering our scenario in which Alice also uses the flipped bases $\bar{b}$.

5.2. THE MODIFIED PROTOCOL. Here, we define the modified protocol and reduce privacy in the original protocol to privacy in the modified protocol. Note that the modified protocol that we will define is not a QKD protocol because Bob does not learn the final key. The modified protocol does not accomplish any practical task, but nevertheless a key is defined and kept secret by Alice. Privacy in this modified protocol means privacy of this key. We will show that, for every eavesdropping strategy in the original protocol, there is a corresponding strategy in the modified protocol so that Eve can obtain as much or more information about the key in the modified protocol as in the original protocol. We will then bound Eve's information in the modified protocol.

We first define an intermediary protocol and next the modified protocol. The intermediary protocol is identical to the original protocol except that Bob uses the opposite basis on the untested positions $i \notin R$. The basic idea behind this approach was first explained and used in Mayers [1995]. To be more precise, let

$$\bar{\mathbf{b}}[i] = \begin{cases} \mathbf{b}[i] & \text{if } i \in \mathbf{R} \\ \overline{\mathbf{b}[i]} & \text{if } i \notin \mathbf{R}. \end{cases}$$

In the intermediary protocol, Bob does as before except that he executes the measurement with the string of bases $\bar{b}$ rather than $b$. The same string $b$ is announced so that the key is defined as before by Alice. This description entirely determines the behavior of the protocol as a random experiment. The set $E$ and $T$ are defined as before in terms of $b$, not $\bar{b}$. We will now show the following proposition.

PROPOSITION 4. *If the algorithm used by Eve to cheat in the intermediary protocol is the same as in the original protocol, the distribution of probability of* (**v**, **k**) *will be identical in both protocols.*

PROOF. For precision, the proof will use the POVM formalism and the terminology of Section 3.4, but the intuition in the proof can be understood without making use of this formalism. The idea is to follow the protocol and see that at every step, whether Bob uses the bases $b$ or the bases $\bar{b}$ to measure the photons, the information that Eve has about the string $g$ is the same in both cases. In particular, it is clear that just after the quantum transmission, before Bob measures the photons, the two cases cannot be distinguished. Next, one can argue that, if we assume that the pair $(\mathcal{D}, h[R])$ announced by Bob has the same distribution of probability whether $b$ or $\bar{b}$ is used, then Eve cannot see any difference. She cannot see any difference because only the bases in $\{1, \ldots, N\} - R$ are flipped and only $(\mathcal{D}, h[R])$ is announced.

The proof formalizes this idea in the POVM formalism. We first describe Eve's attack. The overall quantum system is $H^Q = H^A \otimes H^B \otimes H^E$ where $H^A$ is the state space for the photons, $H^B$ is Bob's received system and $H^E$ is an extra system used by Eve. In the honest protocol, we have $H^B = H^A$ because, in our model of communication, the control over the system $H^A$ (Alice's photons) is simply passed to Bob. In the dishonest case, without loss of generality, we can consider that $H^A$ and $H^B$ are different systems. At the beginning, $H^A$ is controlled by Alice whereas $H^B$ and $H^E$ are controlled by Eve. During the quantum transmission, every thing is controlled by Eve. After the quantum transmission, Eve controls $H^E$ (and $H^A$) and the control over $H^B$ is passed to Bob. Without loss of generality, we assume that no information is left in $H^A$ after the quantum transmission.

The initial random classical information in the protocol (see Section 3.4) is $\hat{c} = (a, b, R, g, \hat{K})$, where $\hat{K}$ denotes the random bits that will be used to pick the matrix $K$. The overall quantum measurement outcome (see Section 3.4) is $\hat{q} = (\mathcal{D}, h, j)$. Eve's view $v = (c, q)$ is the deterministic function of $\hat{v} = (\hat{c}, \hat{q})$ defined by

$$c = (a, b, R, g[R], s, K) \qquad \text{and} \qquad q = (\mathcal{D}, h[R], j),$$

where $s = K \bullet g[E]$ in which $E = \Omega - R$. One must see that it is possible to compute $c$ given $\hat{v}$. It is possible to compute $E$ given $\mathcal{D}$ and $\hat{c}$. The details of the computation of $K$ are not given.

Now, we consider the overall sequence of operations executed in the protocol. We will consider Eve's view later. We denote by $U$ the unitary transformation on $H^Q$ executed by Eve during the quantum transmission. We denote by $E^B_{\mathcal{D},h|b}$ Bob's honest (but possibly defective) measurement operator on $H^B$, which depends on $b$. We denote by $E^E_{j|c,\mathcal{D},h[R]}$ Eve's final measurement operator on $H^E$ executed in view of $(c, \mathcal{D}, h[R])$. In accordance with the extended operator formalism described in Section 3.4, the POVM on $H^C \otimes H^Q$ returning $\hat{v} = (\hat{c}, \hat{q})$ has the form

$$E_{\hat{v}} = E_{\hat{c},\hat{q}} = P_{\hat{c}} E_{\mathcal{D},h,j|\hat{c}}, \tag{14}$$

where

$$E_{\mathcal{D},h,j|\hat{c}} = U^\dagger E^E_{j|\hat{c},\mathcal{D},h[R(\hat{c})]} \otimes E^B_{\mathcal{D},h|b(\hat{c})} U. \tag{15}$$

To derive Eq. (15), one must use the collapse operation formalism. In accordance with this formalism, the POVM $E^E_{j|\hat{c},\mathcal{D},h[R(\hat{c})]}$ on $H^E$ can be written in the form

$$E^E_{j|\hat{c},\mathcal{D},h[R(\hat{c})]} = A^{E^\dagger}_{j|\hat{c},\mathcal{D},h[R(\hat{c})]} A^E_{j|\hat{c},\mathcal{D},h[R(\hat{c})]},$$

where $A^E_{j|\hat{c},\mathcal{D},h[R(\hat{c})]}$ is the collapse operation on $H^E$ associated with $E_{j|\hat{c},\mathcal{D},h[R]}$. Similarly, there is a collapse operation $A^B_{\mathcal{D},h|b(\hat{c})}$ on $H^B$ associated with $E^B_{j|\hat{c},\mathcal{D},h|b(\hat{c})}$. In accordance with Section 3.3, the overall collapse operation on the quantum part $H^Q$, including the initial unitary operation $U$, is

$$A_{\mathcal{D},h,j|\hat{c}} = A^E_{j|\hat{c},\mathcal{D},h[R(\hat{c})]} \otimes A^B_{\mathcal{D},h|b(\hat{c})} U.$$

So the overall POVM on $H^Q$ is

$$E_{\mathcal{D},h,j|\hat{c}} = A^\dagger_{\mathcal{D},h,j|\hat{c}} A_{\mathcal{D},h,j|\hat{c}} = U^\dagger E^E_{j|\hat{c},\mathcal{D},h[R(\hat{c})]} \otimes E^B_{\mathcal{D},h|b(\hat{c})} U,$$

which is consistent with (15).

Now, let us consider the POVM for Eve's view $v$. The only components of $\hat{v} = (\hat{c}, \hat{q})$, which is not a function of $v$ are $g[\bar{R}]$, where $\bar{R} = \{1, \ldots, N\} - R$, and $h[\mathcal{D} - R]$, that is, we can write

$$\hat{v} = \hat{v}(g[\bar{R}], h[\mathscr{D} - R], v).$$

Using formula (7) in Section 3.3 for the first equality and $\hat{v} = \hat{v}(g[\bar{R}], h[\mathscr{D} - R], v)$ for the second equality, we obtain

$$E_v = \sum_{\hat{v}|\mathbf{v}(\hat{v}) = v} E_{\hat{v}}$$

$$= \sum_{g[\bar{R}]} \sum_{h[\mathscr{D} - R]} E_{\hat{v}(g[\bar{R}], h[\mathscr{D} - R], v)}$$

We must obtain that the POVM $E_v$ (the sum of $E_{\hat{v}}$ over all $\hat{v} = \hat{v}(h[\mathscr{D} - R], g[\bar{R}], v)$) is the same in both protocols. We recall that the only difference between the protocols is that $b$ is replaced by $\bar{b}$. The sum is over $h[\mathscr{D} - R]$ and $g[\bar{R}]$, but it will be enough to consider the sum over $h[\mathscr{D} - R]$ only. The only terms in $E_{\hat{v}} = E_{\hat{c}, \hat{q}}$, in formula (14) and (15), which depends upon $h[\mathscr{D} - R]$ is $E^B_{\mathscr{D}, h|b(\hat{c})}$, the operation on Bob's system $H^B$. This fact corresponds to our intuition that what is going on before Bob measures the photons cannot make any difference. Therefore, we must consider the sum

$$\sum_{h[\mathscr{D} - R]} E^B_{\mathscr{D}, h|b},$$

and show that it is the same if we replace $b$ by $\bar{b}$. (In the above sum and from there on, we will not write explicitly the dependence on $\hat{c}$.) To better interpret this sum, note that $h$ is uniquely determined by $\mathscr{D}$, $h[R]$ and $h[\mathscr{D} - R]$, that is, we can replace $E^B_{\mathscr{D}, h|b}$ by $E^B_{\mathscr{D}, h[R], h[\mathscr{D} - R]|b}$. Now, in accordance with formula (7), we have that this sum is nothing else than $E^B_{\mathscr{D}, h[R]|b}$, the POVM associated with the partial outcome $(\mathscr{D}, h[R])$ given $b$, that is,

$$E^B_{\mathscr{D}, h[R]|b} \overset{def}{=} \sum_{h[\mathscr{D} - R]} E^B_{\mathscr{D}, h[R], h[\mathscr{D} - R]|b}.$$

The operator $E^B_{\mathscr{D}, h[R]|b}$ implicitly depends on $R$ because the choice of the set $R$ is part of its formal definition.[3] It is a fact of linear algebra that the requirement that $E^B_{\mathscr{D}, h[R]|b}$ is the same as $E^B_{\mathscr{D}, h[R]|\bar{b}}$ is equivalent to our assumption that, for every state in $H^B$, the distribution of probability of $(\mathscr{D}, h[R])$ is the same whether $b$ or $\bar{b}$ is used to measure the photons. We have obtained under this assumption that Eve's information is the same in the intermediary protocol as in the original protocol. This concludes the proof. $\square$

Now we use the intermediary protocol to describe the modified protocol. In the intermediary protocol, one can assume that there is a box on Bob's side that computes $(R, b, \bar{b})$. This box secretely unveils $\bar{b}$ to Bob at the beginning so that Bob can execute his measurements in these bases. Just before the test, this box publicly announces $R$ and $b$ so that Bob can execute the test with Alice as in the original protocol. This alternative description of the intermediary protocol makes no difference at all for Eve since she does not care about who computes $(R, b,$ $\bar{b}$) and who makes the announcements, as long as the distribution of $(R, b, \bar{b})$, the measurements and the announcements are the same.

---

[3] This is not the same thing as saying that the measurement itself is executed in view of $R$ in the protocol. The dependence on $R$ is OK because $R$ is in $v$ and we are interested in $E_v$.

The modified protocol is like this intermediary protocol except that (1) Bob publicly announces $\bar{b}$ at the beginning, (2) Bob publicly announces $h$ just before the announcement of $R$ by the box, (3) Alice publicly announces $g[\bar{E}]$ where $\bar{E} = \{1, \dots, N\} - E$ after the announcement of $E$ (i.e., of $b$, $R$ and $a$), and (4) Eve can corrupt Bob, but not the box. It is not hard to see that, after each modification, Eve can only have more information or power than she had previously, but even so we shall bound the total information available to Eve in the modified protocol. We gain the advantage that Eve and Bob become like a single participant called Eve–Bob who can use any measurement s/he wants to learn about the key. The new situation, which contains only two participants, is much simpler. The first participant, Alice, sends photons that encode a key, and the second participant, Eve–Bob, tries to find out what is this key via an appropriate measurement. The constraint on Eve–Bob is that $h$ and $\bar{b}$ must be announced before the tested positions $R$ and the string of bases $a$ are known.

5.3. THE MODIFIED PROTOCOL IN OUR MODEL. Essentially, in this section, we apply the basic mechanisms of Section 3.4 to the modified protocol and then provide basic formula that will be useful later in our proof. Let $H^A$ be Alice's original system. In principle, when Eve–Bob receives control over $H^A$ s/he is free to use an extra system $H^B = H^E$. However, Eve–Bob's system $H^B = H^E$ can be considered as an auxiliary system used by Eve–Bob to execute the most general POVM [Peres 1993]. We already use the POVM formalism to describe Eve–Bob's measurement, so without loss of generality, we do not need the extra system $H^B = H^E$. This extra system is implicit in the POVM formalism.

The overall measurement outcome is $\hat{q} = (\mathcal{D}, h, j)$. We have that $\hat{c} = (\bar{b}, a, R, g, \hat{K})$ where, $\hat{K}$ represents the random bits that will be used to generate $K$ (and possibly $F$ if we use a random error-correcting code). The string $b$ is a function of $R$ and $\bar{b}$ and so is not included in $\hat{c}$ (but there would be no harm to include it). The string of classical announcements received by Eve–Bob is $c = (\bar{b}, a, R, E, g[\bar{E}], K, F, s)$. Eve–Bob's quantum outcome is $q = \hat{q} = (\mathcal{D}, h, j)$ where $h$ is the outcome of the measurement executed by Eve–Bob on $H^A$ to pass the test and $j$ is the outcome of the measurement executed by Eve–Bob on the residual system after the first measurement. Eve–Bob's view is $v = (c, q)$.

It is not hard to see that $v$ respects conditions C1 and C2. Without loss of generality, we conservatively consider that the operator $E_{q|c}$ on $H^A$ has rank one, that is, $E_{q|c} = |\phi_{c,q}\rangle\langle\phi_{c,q}|$ for some nonnormalized state $|\phi_{c,q}\rangle$. The initial random state is $|\hat{c}\rangle^C |\Psi(g, a)\rangle$ with probability $p(\hat{c})$ (which probability we will not need to compute). The corresponding density matrix is denoted $\rho$. We have $E_v = P_{c|q} E_{q|c}$. We denote by $\mathrm{Tr}_A$ the trace over $H^A$.

Our first basic formula is for $p(v)$. Using Proposition 2, we obtain

$$p(v) = p(c:q)\mathrm{Tr}_A(E_{q|c}\rho_{c|q}). \tag{16}$$

We compute $\rho_{c|q}$ as explained in Section 3.4 (see remark after Proposition 2). The density matrix $\rho_{c|q}$ is the density matrix that is obtained when we prepare $|\Psi(\hat{c})\rangle$ uniformly at random and only keep the states with $\mathbf{c}_q = c$. We recall that $\mathbf{c}_q \stackrel{def}{=} \mathbf{c}(\hat{c}, q)$. We think of $\hat{c}$ as the outcome of a random experiment on which we define the random variable $\mathbf{c}_q$ (with parameter $q$). In our protocol, we have $\Psi(\hat{c}) = \Psi(g, a)$ and, for given $c$ and $q$, one can check that the constraint $\mathbf{c}_q = c = (\bar{b}, a, R, \ldots, s)$ on $(\mathbf{g}, \mathbf{a})$ corresponds to the three constraints $\mathbf{a} = a$, $\mathbf{g}[\bar{E}] = g[\bar{E}]$ and $F \bullet \mathbf{g}[E] = s$, where $\bar{E} = \{1, \ldots, N\} - E$. Note that, for given $c$ and $q$, the set $E$ and the matrix $F$ are uniquely determined so that $E$ and $F$ can be interpreted as fixed parameters in the above constraints. Let

$$C_s = \{\alpha \in \{0, 1\}^E | F \bullet \alpha = s\}.$$

We obtain that $\rho_{c|q}$ is the product $|\Psi_{\bar{E}}\rangle\langle\Psi_{\bar{E}}| \otimes \tilde{\rho}_s$, where

$$|\Psi_{\bar{E}}\rangle = |\Psi(g[\bar{E}], a[\bar{E}])\rangle \tag{17}$$

is the pure state for the photons in $\bar{E} = \{1, \ldots, N\} - E$ and $\tilde{\rho}_s$ correspond to a uniform distribution over the states $|\Psi(\alpha, a[E])\rangle$ with $\alpha \in C_s$:

$$\tilde{\rho}_s = |C_s|^{-1} \sum_{\alpha \in C_s} |\Psi(\alpha, a[E])\rangle\langle\Psi(\alpha, a[E])|. \tag{18}$$

The state

$$\tilde{\phi}_{c,q} = \langle \underbrace{\Psi_{\bar{E}}}_{\text{On } \bar{E}} | \quad \underbrace{\phi_{c,q}}_{\text{On}\{1,\ldots,N\}} \tag{19}$$

appears naturally in the computation of $p(v)$ because $\rho_{c|q}$ corresponds to the pure state $\Psi_{\bar{E}}$ on $\bar{E}$ and, we recall, $E_{q|c} = |\phi_{c,q}\rangle\langle\phi_{c,q}|$. The component in $\bar{E}$ of $\phi_{c,q}$ is used in the inner product with $\Psi_{\bar{E}}$ so that $\tilde{\phi}_{c,q}$ is the residual state for the photons in $E$ (see Section 3.3). We obtain

$$\mathrm{Tr}_A(\mathrm{E}_{\mathrm{q|c}}\rho_{\mathrm{c|q}}) = \langle\phi_{\mathrm{c,q}}|\rho_{\mathrm{c|q}}|\phi_{\mathrm{c,q}}\rangle$$
$$= \underbrace{\langle\tilde{\phi}_{\mathrm{c,q}}|\tilde{\rho}_s|\tilde{\phi}_{\mathrm{c,q}}\rangle}_{\text{On } E}. \tag{20}$$

So, using (16) and (20), we obtain

$$p(v) = p(c : q)\langle\tilde{\phi}_{c,q}|\tilde{\rho}_s|\tilde{\phi}_{c,q}\rangle. \tag{21}$$

We now prove a generalization of (20). Let $\Pi$ and $\Pi'$ be any two operators on the state space for the photons in $E$. The generalization of (20) is

$$\mathrm{Tr}_A(E_{q|c}\Pi\rho_{c|q}\Pi') = \langle\tilde{\phi}_{c,q}|\Pi\tilde{\rho}_s\Pi'|\tilde{\phi}_{c,q}\rangle. \tag{22}$$

We recall that $\rho_{c|q}$ corresponds to the pure state $\Psi_{\bar{E}}$ on $\bar{E}$ (and to $\tilde{\rho}_s$ on $E$), and $E_{c|q} = |\phi_{c,q}\rangle\langle\phi_{c,q}|$. Also, by hypothesis, $\Pi$ and $\Pi'$ are only defined on $E$. Therefore, the inner product between $\Psi_{\bar{E}}$ and $\phi_{c,q}$ will return the residual state on $E$ as in (21). So we have obtained (22).

It is easy to compute $p(k, v)$ using the same technique. The only difference is that $k$ is added to the view. It is as if the new syndrome was $(s, k)$ rather than $s$ only. We obtain

$$p(k, v) = p(c, k : q)\langle\tilde{\phi}_{c,q}|\tilde{\rho}_{s,k}|\tilde{\phi}_{c,q}\rangle. \tag{23}$$

The normalized density matrix $\bar{\rho}_{s,k}$ is defined as $\bar{\rho}_s$ via (18) except that instead of $C_s$ we use

$$C_{s,k} = \{\alpha \in \{0, 1\}^E | F \bullet \alpha = s \wedge K \bullet \alpha = k\}.$$

In accordance with Section 3.4 (see remark after Proposition 2), $p(c, k : q)$ in (23) is the probability that $(\mathbf{c}, \mathbf{k})_q = (c, k)$ where $(\mathbf{c}, \mathbf{k})_q$ is a function of $\hat{c}$ which is defined in the protocol when $q = (\mathcal{D}, h, j)$ is fixed. We can take the point of view that $(\mathbf{c}, \mathbf{k})_q$ is a random variable defined on $\hat{c}$. For a given $q$, the value of this random variable can be computed by first obtaining $c = \mathbf{c}(\hat{c}, q)$ and then $k$ using $k = K \bullet g[E]$. By definition, the probability of obtaining $c$ in this computation is $p(c : q)$. The probability of every $k$ is $2^{-m}$ independently of the view $v = (c, q)$. So, we have

$$p(c, k : q) = 2^{-m}p(c : q) \tag{24}$$

5.4. PRIVACY IN THE MODIFIED PROTOCOL. In this section, we prove privacy in the modified protocol. The proof follows the work of Mayers [1995; 1996], Mayers and Salvail [1994], and Yao [1995]. There are three important lemmas in the proof: Lemmas 5, 6, and 7. A variation on Lemma 5 for the case $r = 0$ and $m = 1$ and a statement close to Lemma 6 (also for the case $r = 0$ and $m = 1$) was provided in Yao [1995] in the context of a different cryptographic application called quantum oblivious transfer. No variation on Lemma 7 was proven before.

In accordance with Section 2, it is sufficient to obtain that $\mathcal{P} \Rightarrow_\xi \mathcal{N}_\sigma$ where $\mathcal{P} \stackrel{def}{=} \mathcal{P}_T \wedge \mathcal{X}'_2 \wedge \mathcal{X}_3$ and both $\xi$ and $\sigma$ must be exponentially small. The proof breaks the (probabilistic) implication from $\mathcal{P} \Rightarrow_\xi \mathcal{N}_\sigma$ in two implications, $\mathcal{P}_T \Rightarrow_\gamma \mathcal{S}$ and $(\mathcal{S} \wedge \mathcal{X}'_2 \wedge \mathcal{X}_3) \Rightarrow_{\lambda + \eta + 2\sqrt{2\eta}} \mathcal{N}_\sigma$, where $\mathcal{S}$ is called the small sphere property and $\xi = \gamma + \lambda + \eta + 2\sqrt{2\eta}$. The small sphere property as well as $\gamma$, $\lambda$, $\sigma$ and $\eta$ will be defined later.

5.4.1. *Some Intuitions*. In the modified protocol, Bob's bases are flipped. We already explained (see Section 5.1) the intuition that in the proof we need to flip Alice's bases as well. Here, we provide some intuitions from the point of view of Eve–Bob, that is, we start with the modified protocol. We will obtain the same conclusion and more. To go into the essential of the problem we analyze the case where Eve–Bob attacks each photon individually (see Mayers and Salvail [1994] and reference therein for previous analysis). Consider the $i$th photon sent from Alice to Eve–Bob. When Eve–Bob attacks one photon at a time, s/he wants to maximize her information about the bit $g[i]$ for the case where $i \in E$ (i.e., $a[i] \neq \bar{b}[i]$), but at the same time she wants to minimize the probability of creating an error for the case $i \in T$ (i.e., $a[i] = \bar{b}[i]$).

The problem is how to obtain a constraint for the case $i \in E$ using a constraint that results from the case $i \in T$. This issue becomes more important when we consider the most general attack, but it is already not so obvious when we consider the individual attack. In the case of the individual attack, this issue is addressed [Mayers and Salvail 1994] by first obtaining a constraint on Eve–Bob's measurement operators. As a first step, the constraint will only apply to the measurement operators associated with the outcomes $h[i]$. It will not apply to the measurement operators associated with the entire data received by Eve about the bit $g[i]$.

The idea used in Mayers and Salvail [1994] is that this constraint will also be valid in the case $i \in E$, because the measurement operators associated with the outcomes $h[i]$ must be defined by Eve–Bob independently of whether $i$ belongs to $T$ or $E$. This is the case because Bob (in Eve–Bob) immediately notes the value of $h[i]$ when he receives the photons which happens before the announcement of $R$ and $a$. Therefore, this constraint applies to the case $i \in E$ despite the fact it comes from the fictive situation $i \in T$. However, this constraint applies to the outcome $h[i]$ only; it does not directly apply to Eve–Bob's final view after s/he made her final measurement on the photon. This issue can be addressed using the fact that the final measurement operator is a refinement of the incomplete measurement operator associated with $h[i]$ [Mayers and Salvail 1994]. In the case of the most general attack, we will see that a related issue will be addressed in a similar way.

However, one should not look for an exact correspondence between the proof against individual attacks and the proof against all attacks. Here the main ideas that we wanted to emphasize are (1) as in the case of the individual attacks, a fictive test associated with flipped bases on Alice's side will be used to obtain a constraint on Eve–Bob's measurement operator and (2) we need to consider a partial view to connect the fictive test with the real situation.

In this fictive test, Alice prepares the initial random state of the photons in Bob's fixed bases $\bar{b}[i]$, rather than in the random bases $a[i]$. However, $\bar{b}$ is only used for the quantum preparation. The string of bases announced by Alice is $a$ as before, not $\bar{b}$. She uses $(g, \bar{b})$ for the photons. This fictive situation is easier to analyze because (as in the case of the individual attack) we have that Alice and Bob use the same bases to encode and measure the photons respectively.

We will use Lemma 2 to obtain a bound on the number of errors that would be created on $E$ in this fictive situation, the idea being that such a bound will give us a constraint on Eve–Bob's measurement operator, the small sphere property. As in the case of individual attacks, we will have to swap from the fictive situation to the real situation using the fact we consider only a part of Eve's view. The main ingredient that we will use is the fact the initial density matrices for the photons needed to obtain the probability of the partial view $z = (\bar{b}, \mathcal{D}, a, R, h)$ via Proposition 2 are one and the same density matrix, the fully mixed density matrix, in both situations (the fictive and the nonfictive). Note that it is sufficient to swap the bases used in the quantum preparation because Alice's announcements for the bases are the same in the fictive as in the real situation. The bases $\bar{b}$ are only used by Alice in the quantum encoding, not in the classical announcement.

### 5.4.2. *The Strong Small Sphere Property.*

Before analyzing any statement that involves the small sphere property $\mathcal{S}$, we will prove Lemma 5, which is about a closely related property, called the *strong small sphere property*. This lemma will provide some intuition about how the small sphere property $\mathcal{S}$ works and it will directly be used later in the proof of the implication $\mathcal{S} \wedge \mathcal{X}'_2 \wedge \mathcal{X}_3 \Rightarrow_{\lambda + \eta + 2\sqrt{2\eta}} \mathcal{N}_\sigma$. As explained before, the property is given in terms of a fictive preparation where Alice uses Bob's bases $\bar{b}$. For any $\alpha$, we denote by $|\alpha\rangle = \Psi(\alpha, \bar{b}[E])$ the string $\alpha$ encoded in $E$ using Bob's bases $\bar{b}[E]$.

*Definition* 5. Consider any state $\bar{\phi}$ in the state space for the photons in $E$ (not necessarily the state $\bar{\phi}_{c,q}$ given by (19)). We say that $\bar{\phi}$ has the strong small sphere property with radius $d''$ if whenever $\alpha \in \{0, 1\}^E$ does not lie strictly inside the sphere of radius $d''$ around $h[E]$ (in the Hamming distance), we have that $\langle \bar{\phi} | \alpha \rangle = 0$.

*Remark.* In accordance with the basic intuition that was given at the very beginning of the proof, we want to show that Bob receives a lot of information when Alice uses the string of conjugate bases $\bar{a}[E] = \bar{b}[E]$ on $E$. (In the modified protocol, Bob already uses the conjugate bases $\bar{b}[E]$ on $E$). The strong small sphere property (strong ssp) implies that Bob has a lot of information because it puts an upper bound on $d_E(g, h)$, the number of errors in Bob's string $h$ restricted at $E$. The strong ssp says that, given that Alice's initial state is encoded in the string of conjugate bases $\bar{b}$, the outcome associated with $|\bar{\phi}\rangle\langle\bar{\phi}|$ ensure that the number of errors is strictly smaller than $d''$, being implicit here that $|\bar{\phi}\rangle\langle\bar{\phi}|$ is a measurement operator that returns the outcome $h[E]$.

*An Example.* The goal here is to illustrate the strong small sphere property. This example should not be considered as an illustration of the entire proof, only the strong small sphere property is illustrated. We consider a simple kind of attacks where Eve–Bob announced the string of bases $\bar{b}[E] = + + \cdots + +$ on $E$ at the beginning, but Eve–Bob cheated and actually measured in the flipped string of bases $\bar{b}^*[E] = \times \times + \cdots + +$: the bases for the two first positions in $E$ have been flipped with respect to the bases $\bar{b}[E]$. (Eve–Bob can obtain such a situation with a significant probability by flipping few bases at random.) Let us assume that the outcome on $E$ is $h[E] = 00 \cdots 0$. We will see that the state $\Psi(h[E], \bar{b}^*[E])$ has the strong small sphere property with radius 3. The associated "bra" operation is

$$\langle \Psi(h[E], \bar{b}^*[E]) |$$
$$= 1/2(\langle 000 \cdots 0| + \langle 010 \cdots 0| + \langle 100 \cdots 0| + \langle 110 \cdots 0|).$$

There are only four strings $\alpha \in \{0, 1\}^E$ on $E$ such that

$$|\langle \Psi(h[E], \bar{b}^*[E]) | \alpha \rangle| \neq 0.$$

These are the four strings that label the four components of $\langle \Psi(h[E], \bar{b}^*[E])|$. These four strings lie strictly inside a sphere of radius 3 around $h[E]$. Therefore, the state $\Psi(h[E], \bar{b}^*[E])$ has the strong small sphere property with radius 3. $\square$

The strong small sphere property is too strong to be a property of the actual collapse operation executed by Eve–Bob on the photons in $E$. This property cannot be obtained, not even probabilistically. It corresponds to the ideal requirement that the test on $E$ passes with probability exactly 1 given that this collapse operation occurred. Nevertheless, it will be useful in the proof to first consider this ideal situation. The next lemma says that if a state $|\phi\rangle$ has the strong small sphere property then the associated collapse operation provides no information at all about the final key. This lemma combines together privacy amplification and the complementary principle in an intricated manner. The complementary principle is used in the following sense that the strong small sphere property on $|\phi\rangle$ says that the associated collapse operation provides faithful information about Alice's string $g[E]$ if Alice uses the flipped bases $\bar{b}$. Privacy amplification is used because we directly consider the density matrix associated with the final key. We emphasize that the approach in which one first obtains a bound on some kind of information (such as the collision information) about Alice's raw key $g[E]$ and then separately use standard privacy amplification techniques [Bennett et al. 1988] to obtain a much smaller bound on the final key didn't succeed thus far in quantum cryptography.

LEMMA 5.   *For every key $k \in \{0, 1\}^m$ and syndrome $s \in \{0, 1\}^r$, consider the density matrix*

$$\tilde{\rho}_{s,k} \overset{def}{=} |C_{k,s}|^{-1} \sum_{\alpha \in C_{k,s}} |\Psi(\alpha, a[E])\rangle\langle\Psi(\alpha, a[E])|,$$

*where $C_{s,k}$ is the set of string $\alpha \in \{0, 1\}^D$ consistent with the key $k$ and the syndrome $s$, that is, for which $F \bullet g[E] = s$ and $K \bullet g[E] = k$. Consider any state $\tilde{\phi}$ on the state space for the photons in $E$ (not necessarily the state $\bar{\phi}_{c,q}$). If $\tilde{\phi}$ has the strong small sphere property with radius $d'' \leq d_W/2$, then $\langle\tilde{\phi}|\tilde{\rho}_{s,k}|\tilde{\phi}\rangle$ is independent of $k$.*

Consider a fixed view $v = (c, h, j)$. If $\bar{\phi}_{c,q}$ given by (19) has the strong small sphere property with radius $d''$, we say that the view $v$ has the strong small sphere property with radius $d''$.

COROLLARY 1.   *If $v$ has the strong small sphere property with a radius $d'' \leq d_W/2$, then $v$ is 0-informative about every $k$, that is, $p(k|v) = 2^{-m}$ for every $k$.*

PROOF OF THE COROLLARY.   It will be sufficient to show that $p(k|v)$ is independent of $k$. Using (23), we obtain

$$p(k|v) = p(v)^{-1}p(k, v) = p(v)^{-1}p(c, k:q)\langle\bar{\phi}_{c,q}|\tilde{\rho}_{s,k}|\bar{\phi}_{c,q}\rangle.$$

Using (24), we obtain that $p(c, k:q) = 2^{-m}p(c:q)$ is independent of $k$. Using Lemma 5, we also have that $\langle\bar{\phi}_{c,q}|\tilde{\rho}_{s,k}|\bar{\phi}_{c,q}\rangle$ is independent of $k$. This concludes the proof of the corollary.   □

Unlike Lemma 5, this corollary will not be used in the proof of privacy because a successful test does not imply that $v$ has the strong small sphere property, not even probabilistically. This corollary is only provided to support the intuition. It says that something like the strong small sphere property is desired for privacy.

PROOF OF LEMMA 5.   We first do the case where $r = 0$ (no error-correction), $m = 1$ and the one-row binary matrix $K$ is $[11 \cdots 11]$. In this case, we have $d_W = n_E$. Also, one can easily compute the density matrices $\tilde{\rho}_0$ and $\tilde{\rho}_1$, respectively, associated with Alice's preparation for the photons in $E$ when the key is $k = 0$ and $k = 1$. We recall that, for every $\alpha \in \{0, 1\}^{n_E}$, we defined $|\alpha\rangle \overset{def}{=} \Psi(\alpha, \bar{b}[E])$. One obtains that the matrix $[\Delta\tilde{\rho}]_{\{|\alpha\rangle\}}$ of $\Delta\tilde{\rho} \overset{def}{=} \tilde{\rho}_0 - \tilde{\rho}_1$ in Bob's basis $\{|\alpha\rangle\} \overset{def}{=} \{|\alpha\rangle | \alpha \in \{0, 1\}^E\}$ is

$$[\Delta \tilde{\rho}]_{\{|\alpha\rangle\}} = 2^{1-n_E} \begin{pmatrix} \begin{array}{cc|cc} 0 & 1 & & \\ 1 & 0 & & \\ \hline & & \cdot & \\ & & & \cdot \\ \hline & & & \\ \end{array} \begin{array}{cc} & \\ & \\ \hline 0 & 1 \\ 1 & 0 \end{array} \end{pmatrix}$$

in which there are 0 everywhere except when indicated otherwise by the dotted line. The indices $\alpha \in \{0,1\}^E$ for the rows in the matrix are ordered in such a way that any two indices $\alpha_1$, $\alpha_2 \in \{0,1\}^E$ that are at maximal Hamming distance $n_E$ are always adjacent, and the same ordering is used for the indices $\alpha'$ $\in \{0,1\}^E$ for the columns. The entries in the matrix are $\langle \alpha | \Delta \tilde{\rho} | \alpha' \rangle$, where $\alpha$, $\alpha'$ $\in \{0,1\}^E$. We have that $\langle \alpha | \Delta \tilde{\rho} | \alpha' \rangle = 0$ unless $d(\alpha, \alpha') \geq d_W = n_E$. The matrix $[\Delta \tilde{\rho}]_{\{|\alpha\rangle\}} = [\Delta \tilde{\rho}]_{\{|\alpha\rangle\}}^{\langle n_E\rangle}$ can be obtained using the recurrence formula

$$\Delta \tilde{\rho}^{(n)} = \left(\frac{1}{2}\right) \Delta \tilde{\rho}^{(n-1)} \otimes (\tilde{\rho}_0^{(1)} - \tilde{\rho}_1^{(1)}),$$

which can be obtained with some algebra using the formula

$$\tilde{\rho}_b^{(n)} = \left(\frac{1}{2}\right)[\tilde{\rho}_0^{(n-1)} \otimes \tilde{\rho}_b^{(1)} + \tilde{\rho}_1^{(n-1)} \otimes \tilde{\rho}_{\bar{b}}^{(1)}].$$

Now, we want to show that the probability of $v$ is the same given both density matrices. So, we want to show that $\langle \bar{\phi}_{c,q} | \Delta \tilde{\rho} | \bar{\phi}_{c,q} \rangle = 0$. We have that

$$\langle \bar{\phi}_{c,q} | \Delta \rho | \bar{\phi}_{c,q} \rangle = \sum_{\alpha,\alpha'} \langle \bar{\phi}_{c,q} | \alpha \rangle \langle \alpha | \Delta \tilde{\rho} | \alpha' \rangle \langle \alpha' | \bar{\phi}_{c,q} \rangle.$$

We show, in two cases, that every term in the sum is 0.

*Case* 1. If $d(\alpha, \alpha') \geq d_W = n_E$, then, because $\bar{\phi}_{c,q}$ has the strong small sphere property with radius $d_W/2$, either $\langle \bar{\phi}_{c,q} | \alpha \rangle = 0$ or $\langle \alpha' | \bar{\phi}_{c,q} \rangle = 0$.

*Case* 2. If $d(\alpha, \alpha') < n_E$, then $\langle \alpha | \Delta \rho | \alpha' \rangle = 0$. This concludes the proof for the simple case where $m = 1$ and $r = 0$.

Now we do the proof for the general case where $m$, $r > 0$. The matrix $[\tilde{\rho}_{s,k}]_{\alpha,\alpha'}$ is given by

$$(\tilde{\rho}_{s,k})_{\alpha,\alpha'}$$
$$= 2^{-n_E} \begin{cases} 0 & \text{if } (\alpha \oplus \alpha') \notin C^{\perp}[G] \\ (-1)^{\lambda(\alpha \oplus \alpha')\bullet(s,k)} & \text{otherwise,} \end{cases}$$

where

$$G = \begin{pmatrix} F \\ K \end{pmatrix},$$

$C^{\perp}[G]$ is the code generated by $G$ (see Appendix C) and $\lambda$ is the coordinate function that when evaluated on any string $\alpha \in C^{\perp}[G]$ returns the string coordinate $\lambda(\alpha)$ such that $\lambda(\alpha) \bullet G = \alpha$. The computation is provided in Appendix D. By definition of $d_W$, if the weight of $(\alpha \oplus \alpha')$, which is the same as $d(\alpha, \alpha')$, is strictly smaller than $d_W$, then $\lambda(\alpha \oplus \alpha')$ vanishes in its $K$-section. We obtain that, for $(\alpha, \alpha')$ fixed, the sign of the entry $[\tilde{\rho}_{s,k}]_{\alpha,\alpha'}$ depends only on $s$. Therefore, $d(\alpha, \alpha') < d_W$ implies that $[\Delta\tilde{\rho}]_{\alpha,\alpha'} = [\tilde{\rho}_{s,k}]_{\alpha,\alpha'} - [\tilde{\rho}_{s,k'}]_{\alpha,\alpha'}$ vanishes. The remainder of the proof is identical the proof in the simple case, and this can be easily checked by the reader. $\square$

5.4.3. *The First Implication*.   Here we define and discuss the small sphere property $\mathcal{S}$ and prove the implication $\mathcal{P}_T \Rightarrow_\gamma \mathcal{S}$ (see Lemma 6). For any set of positions $X$ and any integer $d'' \geq 0$, let $\Pi_0[X, d'']$ be the projection on the span of $\{\Psi(\alpha, \tilde{b}) | d_X(\alpha, h) \geq d''\}$ where $d_X(\alpha, h)$ is the Hamming distance between $\alpha$ and $h$ on $X$.

*Definition* 6.   Let $\tilde{\rho}_s$ be defined as in Section 5.3. The view $v$ has the small sphere property $\mathcal{S}$ with radius $d'' > 0$ and precision $\gamma > 0$ if

$$p(c:q)\langle\tilde{\phi}_{c,q}|\Pi_0[E, d'']\tilde{\rho}_s\Pi_0[E, d'']|\tilde{\phi}_{c,q}\rangle \leq \gamma p(v) \tag{25}$$

or equivalently

$$\langle\tilde{\phi}_{c,q}|\Pi_0[E, d'']\tilde{\rho}_s\Pi_0[E, d'']|\tilde{\phi}_{c,q}\rangle \leq \gamma\langle\tilde{\phi}_{c,q}|\tilde{\rho}_s|\tilde{\phi}_{c,q}\rangle. \tag{26}$$

We have already given part of the intuition for the small sphere property when we discussed the strong small sphere property. Now, we explain the connection with this intuition. We recall that the strong small sphere property says that any state encoded in Bob's basis $\tilde{b}[E]$ on or outside the small sphere is rejected by $\tilde{\phi}_{c,q}$. In this way the strong small sphere property says that $v$ provides faithful information about Alice's string $g[E]$ if Alice uses Bob's basis $\tilde{b}[E]$. Lemma 5 is in fact a strong version of the complementary principle because it concludes that the view $v$ provides no information at all about the key encoded in Alice's original bases. The states that are on or outside the small sphere span the space associated with the projection $\Pi_0[E, d'']$. Modulo some small imprecision (that is quantified by a small value $\gamma$), if we expand the density matrix $\rho_s$, we see that the small sphere property expresses a similar requirement as the strong small sphere property, except that this requirement is now expressed in terms of a mixture of states $\Pi_0[E, d'']\Psi(\alpha, a[E])$, $\alpha \in_R C_s$, obtained from Alice's bases $a[E]$, not Bob's bases. The indirect connection with Bob's bases is provided by the projection $\Pi_0[E, d'']$, which is defined in Bob's basis $\tilde{b}[E]$. As we will see, this will be close enough to the strong small sphere property. The (not strong) small sphere property has the technical advantage that it is written in terms of Alice's original preparation: the matrix $\tilde{\rho}_s$. An alternative small sphere property stated in terms of Bob's bases could have been more in accord with the complementary principle but it would have been difficult to use, and even to obtain. It works better to only have an indirect connection with Bob's bases via the projection $\Pi_0[E, d'']$.

Note that in the case $r = 0$ (no error-correction), $\tilde{\rho}_s$ is proportional to the identity matrix and therefore, when $r = 0$, the small sphere property with radius $d''$ is equivalent to

$$\|\Pi_0[E, d'']\tilde{\phi}_{c,q}\|^2 \leq \gamma\|\tilde{\phi}_{c,q}\|^2.$$

In the exact case $\gamma = 0$, the last inequality is the strong small sphere property on $v$. The basic idea for the (strong) small sphere property was first published in Yao [1995] in the context of the security of QOT. Historically, the small sphere property $\mathscr{S}$ was obtained by trying to prove an implication of the form $\mathscr{P}_T \Rightarrow_\gamma \mathscr{S}$ where $\mathscr{S}$ is as close as possible to the strong small sphere property on $v$ [Mayers 1996].

Some brief recapitulation. The next lemma, Lemma 6, says how the test probabilistically implies the small sphere property $\mathscr{S}$. We just explained that this small sphere property is the kind of hypothesis that is required to apply the complementary principle. This complementary principle will be expressed in Lemma 7, the small sphere property $\mathscr{S}$ being the required hypothesis. The proof of Lemma 7 makes use of Lemma 5, its strong version in which the strong small sphere property is the required hypothesis.

LEMMA 6. *Let $\gamma = \mu(\epsilon, n_\Omega^{min})^{1/2}$ where the function $\mu$ is defined as in Theorem 1. We have $\mathscr{P}_T \Rightarrow_\gamma \mathscr{S}$, where $\mathscr{S}$ is the small sphere property with radius $d'' = d_+(\epsilon)$ and precision $\gamma$.*

PROOF OF LEMMA 6. Let $\mathscr{P}_E$ be the event that the number of errors in $E$ is smaller than $d_+(\epsilon)$. All the ingredients that are mentioned above suggest that in the proof we must use the probabilistic implication

$$\mathscr{P}_T \Rightarrow_{\mu(\epsilon, n_\Omega)} \mathscr{P}_E \tag{27}$$

in which Alice uses the basis $\{\Psi(g, \bar{b}) | g \in \{0, 1\}^N\}$ rather than the basis $\{\Psi(g, a) | g \in \{0, 1\}^N\}$. (In this fictive test, we flip Alice's bases when $a[i] \neq \bar{b}[i]$.)

We first explain how Lemma 2 applies to this fictive situation. The important ingredient in this lemma is the distribution of probability of the two sets $E$ and $T$ for a fixed error string $g[D] \oplus h[D]$. The fictive preparation is an encoding in the bases $\bar{b}$ which is fixed. Eve–Bob determines and executes the measurement without knowing $a$ and $R$. We fix $\mathscr{D}$, $\Omega$ and the error string on $\Omega$, but keep $a$ and $R$ random. The set $\Omega$ will play the role of the set $D$ in the lemma. It will be sufficient that the distribution of probability for $E = \Omega - R$ and $T = \Omega \cap R$ is as required in the lemma. This is exactly the case because every position in $\Omega$ is put in $R$ with probability $p_T$. One might find strange that it seems that it is not required that the bases $a$ are chosen at random. In fact, they also have to be chosen at random because $(\bar{b}, \Omega, a)$ uniquely determines $R$. So we have obtained that Lemma 2 applies with $D = \Omega$ and thus we have obtained inequality (27).

Note that it is implicit in (27) that we consider the context where $\Omega$ is fixed because the statement in Lemma 2, in particular the definition of $\mu(\beta, n_D)$, assume that the set $D$ is fixed (and $\Omega$ plays the role of $D$). We define $\mathscr{X} : \Omega = \Omega$. In fact, instead of (27), what we have is $\mathscr{P}_T \Rightarrow_{\mu(\epsilon, n_\Omega) | \mathscr{X}} \mathscr{P}_E$. Note that $\mathscr{P}_T \Rightarrow_{\mu | \mathscr{X}} \mathscr{P}_E$ is equivalent to $\mathscr{P}_T \wedge \mathscr{X} \Rightarrow_{\mu \times \Pr(\mathscr{X})} \mathscr{P}_E$. Let $\mathscr{P}'_T = \mathscr{P}_T \wedge \mathscr{X}$. The starting point is in fact

$$\mathscr{P}'_T \Rightarrow_{\mu(\epsilon, n_\Omega) \times \Pr(\mathscr{X})} \mathscr{P}_E. \tag{28}$$

Note that Eve–Bob's attack is uniquely determined by the POVM $E_{q|c}$. We must show that the probabilistic implication $\mathscr{P}_T \Rightarrow_\gamma \mathscr{S}$ hold for all POVM $E_{q|c}$ that corresponds to an attack in the real protocol where Alice uses the basis $\{\Psi(g, a)|g \in \{0, 1\}^N\}$ (not the basis $\{\Psi(g, \bar{b})|g \in \{0, 1\}^N\}$). Let us consider any such POVM in the real protocol (this POVM will not be further restricted so that the proof will apply to any such POVM). The interesting point is that any such POVM $E_{q|c}$ still corresponds to an attack in the protocol even if Alice uses the basis $\{\Psi(g, \bar{b})|g \in \{0, 1\}^N\}$. In fact, if we don't tell Eve–Bob that a different preparation was used for the photons, then the same POVM will be executed. Therefore, because the probabilistic implication (28) is valid against all attacks, this probabilistic implication must apply to the POVM $E_{q|c}$. Here, we translate (28) in terms of density matrices and projections operators. Let $\rho_{\bar{b}}$ be the density matrix that corresponds to the state $\Psi(g, \bar{b})$ with probability $2^{-N}$. This is the fully mixed density matrix $2^{-N}\mathbf{I}$. Let $\hat{\Pi}_1 = \mathbf{I} - \Pi_0[T, d]$ and $\hat{\Pi}_0 = \Pi[E, d_+(\epsilon)]$. These two projections are respectively associated with the success of the test on $T$ and the failure of the test on $E$ (when Alice uses the basis $\{\Psi(g, \bar{b})|g \in \{0, 1\}^N\}$). Let $z$ be the partial view $(\bar{b}, \mathcal{D}, h, a, R)$. Note that $z$ contains all the necessary information to define the projections $\hat{\Pi}_0$ and $\hat{\Pi}_1$. We obtain

$$\Pr(\mathscr{P}'_T \wedge \bar{\mathscr{P}}_E)$$

$$= \sum_{(\bar{b}, a, R, \mathcal{D}, h)|\mathscr{X}} p(\bar{b}, a, R)\mathrm{Tr}_{\mathcal{A}}(E_{\mathcal{D}, h|a, R}\hat{\Pi}_1\hat{\Pi}_0\rho_{\bar{b}}\hat{\Pi}_0\hat{\Pi}_1)$$

in which $\mathscr{X}$ is the constraint $\mathbf{\Omega}(z) = \Omega$ on $z = (\bar{b}, \mathcal{D}, h, a, R)$. This last equality can easily be verified by expanding the density matrix $\rho_{\bar{b}}$ as a sum over state and noting that each term in the sum is annihilated by $\hat{\Pi}_1\hat{\Pi}_0$ if and only if the corresponding state is not consistent with $\mathscr{P}_T \wedge \bar{\mathscr{P}}_E$. Formula (28), which corresponds to the inequality $\Pr(\mathscr{P}_T \wedge \bar{\mathscr{P}}_E) \leq \mu(\epsilon, n_\Omega)\Pr(\mathscr{X})$, becomes

$$\sum_{(\bar{b}, a, R, \mathcal{D}, h)|\mathscr{X}} p(\bar{b}, a, R)\mathrm{Tr}_{\mathcal{A}}(E_{\mathcal{D}, h|\bar{b}, a, R}\hat{\Pi}_1\hat{\Pi}_0\rho_{\bar{b}}\hat{\Pi}_0\hat{\Pi}_1)$$

$$\leq \mu(\epsilon, n_\Omega)\Pr(\mathscr{X}). \tag{29}$$

Note that we have chosen the partial view $z$ so that it contains no classical announcement about $g$. This is a key ingredient which implies that the density matrix for the photons given the classical part $(\bar{b}, a, R)$ in $z$ is $2^{-N}\mathbf{I}$. We can replace $\rho_{\bar{b}}$ by $\rho_a$ because these two density matrices are one and the same density matrix, the fully mixed density matrix. (Note that the string $g$ is independent of $\mathbf{\Omega}$ so that $g$ is still uniformly distributed in the context $\mathscr{X}: \mathbf{\Omega} = \Omega$.) The reader can easily compute this density matrix because it is a product of two dimensional density matrices, each of them corresponding to a state uniformly picked at random in the corresponding basis. By definition a uniform mixture of state in a given basis corresponds to the matrix $\rho^{(2)} = (1/2)|0\rangle\langle 0| + (1/2)|1\rangle\langle 1|$ in that basis. The reader can easily check that, keeping the same representational basis for the density matrix, but considering a uniform mixture of states in the conjugate basis (or any other basis) one obtains the same density matrix. The density matrix $\rho_a$ corresponds to Alice's preparation in the real protocol. We have indirectly obtained a property on the view $z$ for most $z$ since we have an upper bound on the sum in (29). The idea is that it is not possible that many terms in the sum are large when the sum is small. However, what we need is a property on the final view $v$. So, we need to replace the sum over $z$ by a sum over $v$. In this way, we will obtain a property on most $v$. To pass from a sum over $z$ to a sum over $v$, we need to use the extended operator formalism. We will use (12) with the view

$$z = ((\underbrace{\bar{b}, a, R}_{c}), \underbrace{(\mathcal{D}, h)}_{q}))$$

in which $c$ and $q$ refer to the generic notions of Section 3.4, not to $c$ and $q$ defined in this section. Working on the left-hand side of (29), we obtain

$$\sum_{z|\mathcal{X}} \mathrm{Tr}_A(E_z \hat{\Pi}_1 \hat{\Pi}_0 \rho_a \hat{\Pi}_0 \hat{\Pi}_1) \leq \mu(\epsilon, n_\Omega) \mathrm{Pr}(\mathcal{X}).$$

Now, we can use the formula $E_z = \sum_{v|\mathbf{z}=z} E_v$ to obtain

$$\sum_{v|\mathcal{X}} \mathrm{Tr}_A(E_v \hat{\Pi}_1 \hat{\Pi}_0 \rho_a \hat{\Pi}_0 \hat{\Pi}_1) \leq \mu(\epsilon, n_\Omega) \mathrm{Pr}(\mathcal{X}). \tag{30}$$

To return to the standard formalism, we apply again (12), but this time with the complete view $v = (c, q)$. We obtain

$$\sum_{(c,q)|\mathcal{X}} p(c:q) \mathrm{Tr}_A(E_{q|c} \hat{\Pi}_1 \hat{\Pi}_0 \rho_{c|q} \hat{\Pi}_0 \hat{\Pi}_1) \leq \mu(\epsilon, n_\Omega) \mathrm{Pr}(\mathcal{X}).$$

We recall that a mixture associated with $\rho_{c|q}$ corresponds to the pure state $\Psi(g[T], a[T])$ on $T$. The projection $\hat{\Pi}_1$, which is associated with the success of $\mathcal{P}_T$, is also defined in the bases $a[T]$ on $T$. Therefore, every term in the sum with $v \notin \mathcal{P}_T$ vanishes. For the other terms we have

$$\mathrm{Tr}_A(E_{q|c} \hat{\Pi}_1 \hat{\Pi}_0 \rho_{c|q} \hat{\Pi}_0 \hat{\Pi}_1) = \mathrm{Tr}_A(E_{q|c} \hat{\Pi}_0 \rho_{c|q} \hat{\Pi}_0).$$

So we have

$$\sum_{(c,q)|\mathcal{P}_T} p(c:q) \mathrm{Tr}_A(E_{q|c} \hat{\Pi}_0 \rho_{c|q} \hat{\Pi}_0) \leq \mu(\epsilon, n_\Omega) \mathrm{Pr}(\mathcal{X}).$$

At this point, the following lemma must be used. It says that it is not possible that a sum with positive terms contains many large terms when the sum is small.

PROPOSITION 5. *Consider any* $\mu > 0$ *and let* $p(y)$ *be any distribution of probability on a set* $Y$. *Let* $a_y, y \in Y$, *be positive real numbers such that* $\sum_{y \in Y} a_y \leq \mu$. *Consider any positive number* $q > 0$. *We have that*

$$Pr(a_y \geq q\mu p(y)) \leq \frac{1}{q},$$

*that is, except with a probability smaller than* $1/q$, $a_y < q\mu p(y)$.

PROOF OF PROPOSITION 5. Let us denote

$$S = \{y \in Y | a_y \geq q\mu p(y)\}.$$

Assume to the contrary that $\Pr(S) > 1/q$. We obtain

$$\sum_y a_y = \sum_{y \in S} a_y + \sum_{y \notin S} a_y$$

$$\geq \sum_{y \in S} a_y \geq q\mu \sum_{y \in S} p(y) > \left(\frac{1}{q}\right) q\mu = \mu,$$

which contradicts the hypothesis of the lemma. $\square$

Let $\gamma = \mu(\epsilon, n_\Omega)^{1/2}$. At this stage $\gamma$ depends on $n_\Omega$, not on $n_\Omega^{\min}$, but we will take care of this issue later. We will use Proposition 5 with $q = \Pr(\mathscr{P}'_T)/(\gamma \Pr(\mathscr{X})) = \Pr(\mathscr{P}_T|\mathscr{X})/\gamma$ and $p(y) = p(v)/\Pr(\mathscr{P}'_T)$, that is, $p(y)$ is the probability of $v$ conditioned with the event $\mathscr{P}'_T$. We obtain that, conditioned with the event $\mathscr{P}'_T$, except with probability $\gamma/\Pr(\mathscr{P}_T|\mathscr{X})$, we have

$$p(c:q)\mathrm{Tr}_A(E_{q|c}\hat{\Pi}_0\rho_{c|q}\hat{\Pi}_0) \leq \gamma \times p(v). \tag{31}$$

Using (22), we obtain that (31) is equivalent to

$$p(c:q)\langle \bar{\phi}_{c,q}|\hat{\Pi}_0\tilde{\rho}_s\hat{\Pi}_0|\bar{\phi}_{c,q}\rangle \leq \gamma \times p(v),$$

which is the small sphere property $\mathscr{S}$ with radius $d_+(\epsilon)$ and precision $\gamma$. We have obtained that $\gamma/\Pr(\mathscr{P}_T|\mathscr{X})$ is an upper bound for the probability that the small sphere property $\mathscr{S}$ fails given $\mathscr{P}'_T$. We obtain that

$$\Pr(\mathscr{P}_T \wedge \bar{\mathscr{S}}|\mathscr{X}) = \Pr(\bar{\mathscr{S}}|\mathscr{P}'_T)\Pr(\mathscr{P}_T|\mathscr{X}) \leq \gamma.$$

Thus, we have

$$\Pr(\mathscr{P}_T \wedge \bar{\mathscr{S}}|\mathbf{n}_\Omega = n_\Omega) \leq \gamma$$

or equivalently

$$\mathscr{P}_T \Rightarrow_{\gamma|\mathbf{n}_\Omega = n_\Omega} S.$$

The value $\gamma$ depends on the value $n_\Omega$. We can easily take care of this problem and replace $n_\Omega$ by $n_\Omega^{\min}$ using Proposition 6 in Appendix E (we must think of $S$ as an event-valued function, where $n_\Omega$ in the precision $\gamma$, plays the role of the integer $l$ in Proposition 6). The radius $d_+(\epsilon)$ will still depend upon the random value $n_\Omega$ (but we will take care of this problem in the next lemma). This concludes the proof of $\mathscr{P}_T \Rightarrow_\gamma \mathscr{S}$. $\square$

5.4.4. *The Second Implication.* Here we prove that the small sphere property probabilistically implies that Eve's view $v$ is $\sigma$-informative.

LEMMA 7. *Let $\epsilon > 0$ and $\tau > 0$ be the (nonphysical) parameters chosen in the validation constraint $\mathscr{X}'_2$. Consider any $\gamma > 0$. Let $\lambda = 2^{-\tau n_E^{\min}}$, $\eta = 2\sqrt{\gamma} + \gamma$ and $\sigma = \eta + \sqrt{2\eta}$. Except with probability $\lambda + \eta + 2\sqrt{2\eta}$, if we have the constraint $\mathscr{X}'_2 \wedge \mathscr{X}_3$ and the view $v$ has the small sphere property $\mathscr{S}$ with radius $d_+(\epsilon)$ and precision $\gamma$, then the view $v$ is $\sigma$-informative.*

*Remark.* Later, we will use this lemma with $\gamma$ that was defined in Lemma 6, but the lemma hold for every $\gamma > 0$. Lemma 7 corresponds to the implication $\mathscr{S} \wedge \mathscr{X}_2' \wedge \mathscr{X}_3 \Rightarrow_{\lambda + \eta + 2\sqrt{2\eta}} \mathscr{N}_\sigma$.

PROOF OF LEMMA 7. The basic idea of the proof is simple. Consider again formula (23):

$$p(v, k) = p(c, k : q)\langle \bar{\phi}_{c,q} | \bar{\rho}_{s,k} | \bar{\phi}_{c,q} \rangle.$$

We insert the identity operator $\Pi_0[E, \lfloor d_W/2 \rfloor] + \Pi_1[E, \lfloor d_W/2 \rfloor]$ on both sides of $\bar{\rho}_{s,k}$, and we rewrite the expression using terms that contain only $\Pi_0[E, \lfloor d_W/2 \rfloor]$ or only $\Pi_1[E, \lfloor d_W/2 \rfloor]$, not both projections. The term with the projection $\Pi_1[E, \lfloor d_W/2 \rfloor]$ will be taken care by Lemma 5, because the state $\Pi_1[E, \lfloor d_W/2 \rfloor]|\bar{\phi}_{c,q}\rangle$ has the strong small sphere property with radius $d''$ smaller than $d_W/2$. The other terms will be small because of the definition of the small sphere property.

However, before we do that, we must take care of some technical issue related to the use of $\mathscr{X}_2'$ in place of $\mathscr{X}_2$. We recall that, given that the $m \times n_E$ privacy matrix $K$ is chosen uniformly at random, Lemma 4 tells us that, for any $\tau > 0$, we have

$$\text{True} \Rightarrow_\lambda d_W \geq H^{-1}\left(1 - \frac{r + m}{n_E} - \tau\right)n_E, \tag{32}$$

where $\lambda = 2^{-\tau n_E}$. We recall that $\mathscr{X}_2'$ says

$$H^{-1}\left(1 - \frac{r + m}{n_E} - \tau\right) \times n_E \geq 2d_+(\epsilon) \tag{33}$$

for some fixed $\tau > 0$ and $\epsilon > 0$. By transitivity, the inequalities in (32) and (33) implies $\mathscr{X}_2$ which states $d_W \geq 2d_+(\epsilon)$. Therefore, we have $\mathscr{X}_2' \Rightarrow_\lambda \mathscr{X}_2$ and thus $(\mathscr{S} \wedge \mathscr{X}_2') \Rightarrow_\lambda (\mathscr{S} \wedge \mathscr{X}_2)$. Here, the probability $\lambda$ is conditioned by $n_E$, but we can replace $n_E$ by $n_E^{\min}$ in the definition of $\lambda$ using Proposition 6 in Appendix E, adding the constraint $\mathscr{X}_3$ on the left-hand side. Let $\mathscr{S}'$ be the small sphere property with radius $\lfloor d_W/2 \rfloor$ and the same precision $\gamma$ as for $\mathscr{S}$. We have that $\mathscr{S} \wedge \mathscr{X}_2 \wedge \mathscr{X}_3 \Rightarrow \mathscr{S}'$ because the radius in $\mathscr{S}'$ is greater (a weaker constraint). So it will be sufficient to show $\mathscr{S}' \Rightarrow_{\eta + 2\sqrt{2\eta}} \mathscr{N}_\sigma$, where $\mathscr{S}'$ is the small sphere property with radius $\lfloor d_W/2 \rfloor$ and precision $\gamma$.

We start by finding out what can be obtained nonprobabilistically from $\mathscr{S}'$, keeping in mind that we want to bound $|p(k|v) - 2^{-m}| = p(v)^{-1}|p(v, k) - 2^{-m}p(v)|$ probabilistically. We will start by considering the quantities $p(v, k)$ and $2^{-m}p(v)$ separately. We begin by $p(v, k)$. Using (23), one obtains

$$p(v, k) = p(c, k : q)\langle \bar{\phi}_{c,q} | \bar{\rho}_{s,k} | \bar{\phi}_{c,q} \rangle.$$

Now, let us define $\tilde{\Pi}_0 = \Pi_0[E, \lfloor d_W/2 \rfloor]$ and $\tilde{\Pi}_1 = \mathbf{I} - \tilde{\Pi}_0$. The projections $\tilde{\Pi}_0$ and $\tilde{\Pi}_1$ are respectively associated with the failure and the success of the test on $E$ with a tolerated number of errors $\lfloor d_W/2 \rfloor$. If one puts the identity operator $\mathbf{I} =$

$\check{\Pi}_0 + \check{\Pi}_1$ on both sides of $\tilde{\rho}_{k,s}$, after some algebra, one obtains

$$p(v, k) = 2^{-m} p(c : q)$$

$$[\langle \bar{\phi}_{c,q} | \tilde{\Pi}_1 \tilde{\rho}_{s,k} \tilde{\Pi}_1 | \bar{\phi}_{c,q} \rangle + \langle \bar{\phi}_{c,q} | \tilde{\Pi}_0 \tilde{\rho}_{s,k} | \bar{\phi}_{c,q} \rangle$$

$$+ \langle \bar{\phi}_{c,q} | \tilde{\rho}_{s,k} \tilde{\Pi}_0 | \bar{\phi}_{c,q} \rangle - \langle \bar{\phi}_{c,q} | \tilde{\Pi}_0 \tilde{\rho}_{s,k} \tilde{\Pi}_0 | \bar{\phi}_{c,q} \rangle],$$

where we used (24) to replace $p(c, k : q)$ by $2^{-m} p(c : q)$. Note that $\tilde{\Pi}_1 \bar{\phi}_{c,q}$ is a state that has the strong small sphere property with radius $\lfloor d_W/2 \rfloor$. Therefore, using Lemma 5, one obtains that

$$p = 2^{-m} p(c : q) \langle \bar{\phi}_{c,q} | \tilde{\Pi}_1 \tilde{\rho}_{s,k} \tilde{\Pi}_1 | \bar{\phi}_{c,q} \rangle$$

is independent of $k$. In particular, after a sum over $k$ on both sides, we have

$$2^m p = p(c : q) \langle \bar{\phi}_{c,q} | \tilde{\Pi}_1 \tilde{\rho}_s \tilde{\Pi}_1 | \bar{\phi}_{c,q} \rangle$$

or equivalently

$$p = 2^{-m} p(c : q) \langle \bar{\phi}_{c,q} | \tilde{\Pi}_1 \tilde{\rho}_s \tilde{\Pi}_1 | \bar{\phi}_{c,q} \rangle.$$

We denote $\Delta_{s,k} = |p(v, k) - p|$. Since

$$|\langle \bar{\phi}_{c,q} | \tilde{\Pi}_0 \tilde{\rho}_{s,k} | \bar{\phi}_{c,q} \rangle| = |\langle \bar{\phi}_{c,q} | \tilde{\rho}_{s,k} \tilde{\Pi}_0 | \bar{\phi}_{c,q} \rangle|,$$

we have

$$\Delta_{s,k} \leq 2^{-m} p(c : q)$$

$$\times [2|\langle \bar{\phi}_{c,q} | \tilde{\Pi}_0 \tilde{\rho}_{s,k} | \bar{\phi}_{c,q} \rangle| + |\langle \bar{\phi}_{c,q} | \tilde{\Pi}_0 \tilde{\rho}_{s,k} \tilde{\Pi}_0 | \bar{\phi}_{c,q} \rangle|]. \tag{34}$$

Now we bound the first term in the square bracket. We use

$$|\langle \bar{\phi}_{c,q} | \tilde{\Pi}_0 \tilde{\rho}_{s,k} | \bar{\phi}_{c,q} \rangle| = |\underbrace{\langle \bar{\phi}_{c,q} | \tilde{\Pi}_0 \tilde{\rho}_{s,k}^{1/2}}_{\xi^\dagger} \underbrace{\tilde{\rho}_{s,k}^{1/2} | \bar{\phi}_{c,q} \rangle}_{\mathscr{X}}|.$$

Using Schwartz inequality, $|\xi^\dagger \mathscr{X}| \leq \|\xi\| \times \|\mathscr{X}\|$, one obtains

$$|\langle \bar{\phi}_{c,q} | \tilde{\Pi}_0 \tilde{\rho}_{s,k} | \bar{\phi}_{c,q} \rangle|$$

$$\leq \|\tilde{\rho}_{s,k}^{1/2} \tilde{\Pi}_0 | \bar{\phi}_{c,q} \rangle\| \, \|\tilde{\rho}_{s,k}^{1/2} | \bar{\phi}_{c,q} \rangle\|$$

$$= \langle \bar{\phi}_{c,q} | \tilde{\Pi}_0 \tilde{\rho}_{s,k} \tilde{\Pi}_0 \bar{\phi}_{c,q} \rangle^{1/2} \langle \bar{\phi}_{c,q} | \tilde{\rho}_{s,k} | \bar{\phi}_{c,q} \rangle^{1/2}.$$

Therefore

$$\Delta_{s,k} \leq \quad 2^{-m} p(c : q)$$

$$[2 \langle \bar{\phi}_{c,q} | \tilde{\Pi}_0 \tilde{\rho}_{s,k} \tilde{\Pi}_0 | \bar{\phi}_{c,q} \rangle^{1/2} \langle \bar{\phi}_{c,q} | \tilde{\rho}_{s,k} | \bar{\phi}_{c,q} \rangle^{1/2}$$

$$+ \langle \bar{\phi}_{c,q} | \tilde{\Pi}_0 \tilde{\rho}_{s,k} \tilde{\Pi}_0 | \bar{\phi}_{c,q} \rangle].$$

Now, we define $\Delta_s = |p - 2^{-m} p(v)|$. We have

$$|p(v, k) - 2^{-m}p(v)| \leq \Delta_s + \Delta_{s,k},$$

so that the property $\mathcal{N}_\sigma$ is equivalent to $\Delta_s + \Delta_{s,k} \leq 2^{-m}p(v)\sigma$. Using the same technique that we used for $\Delta_{s,k}$, we obtain

$$\Delta_s \leq 2^{-m}p(c\!:\!q)$$
$$[2\langle\bar{\phi}_{c,q}|\tilde{\Pi}_0\tilde{\rho}_s\tilde{\Pi}_0|\bar{\phi}_{c,q}\rangle^{1/2}\langle\bar{\phi}_{c,q}|\tilde{\rho}_s|\bar{\phi}_{c,q}\rangle^{1/2}$$
$$+ \langle\bar{\phi}_{c,q}|\tilde{\Pi}_0\tilde{\rho}_s\tilde{\Pi}_0|\bar{\phi}_{c,q}\rangle.$$

Using (21), which states $p(c\!:\!q)\langle\bar{\phi}_{c,q}|\tilde{\rho}_s|\bar{\phi}_{c,q}\rangle = p(v)$, and the small sphere property (Definition (6)), which states

$$p(c\!:\!q)\langle\bar{\phi}_{c,q}|\tilde{\Pi}_0\tilde{\rho}_s\tilde{\Pi}_0|\bar{\phi}_{c,q}\rangle \leq \gamma p(v),$$

we obtain

$$\Delta_s \leq 2^{-m}(2\sqrt{\gamma} + \gamma)p(v).$$

Since the expression $(2\sqrt{\gamma} + \gamma)$ will appear often, we denote $\eta = 2\sqrt{\gamma} + \gamma$. We obtain

$$\Delta_s \leq 2^{-m}\eta p(v). \tag{35}$$

Now, to bound $\Delta_{s,k}$, it will be useful to define

$$a_0(v, k) = 2^{-m}p(c\!:\!q)\langle\bar{\phi}_{c,q}|\tilde{\Pi}_0\tilde{\rho}_{s,k}\tilde{\Pi}_0|\bar{\phi}_{c,q}\rangle$$

and

$$a_1(v, k) = 2^{-m}p(c\!:\!q)\langle\bar{\phi}_{c,q}|\tilde{\rho}_{s,k}|\bar{\phi}_{c,q}\rangle.$$

We obtain

$$\Delta_{s,k} \leq 2\sqrt{a_1(k, v)a_0(k, v)} + a_0(k, v). \tag{36}$$

Thus far, we have shown

$$\mathcal{S}' \Rightarrow \Delta_s \leq 2^{-m}\eta p(v)$$
$$\wedge \Delta_{s,k}$$
$$\leq 2\sqrt{a_1(k, v)a_0(k, v)} + a_0(k, v).$$

In the remainder of the proof, we will define an event $\mathcal{M}_q$ with a parameter $q > 0$ so that $\mathcal{M}_q \Rightarrow \mathcal{N}_{\eta+q\eta}$, and, for $q = \sqrt{2/\eta}$, we will show $\mathcal{S}' \Rightarrow_{\eta+2\sqrt{2\eta}} \mathcal{M}_q$. This will be sufficient to conclude $\mathcal{S}' \Rightarrow_{\eta+2\sqrt{2\eta}} \mathcal{N}_\sigma$. We recall that $\sigma = \eta + \sqrt{2\eta}$.

Now, we define $\mathcal{M}_q$. To obtain $\mathcal{N}_\sigma$ from $\mathcal{M}_q$, the event $\mathcal{M}_q$ must provide upper bounds for $a_0(k, v)$ and $a_1(k, v)$.[4] For every $q > 0$, we define the events

$$\mathcal{M}_q^{(0)} : a_0(k, v) \le q\gamma p(v)2^{-m},$$

and

$$\mathcal{M}_q^{(1)} : a_1(k, v) \le qp(v)2^{-m},$$

and $\mathcal{M}_q = \mathcal{M}_q^{(0)} \wedge \mathcal{M}_q^{(1)}$. It is not hard to see that

$$\mathcal{M}_q \Rightarrow \Delta_{s,k} \le 2\sqrt{q\gamma p(v)2^{-m}qp(v)2^{-m}} + q\gamma p(v)2^{-m}$$

$$= qp(v)2^{-m}(2\sqrt{\gamma} + \gamma)$$

$$= qp(v)2^{-m}\eta.$$

Therefore,

$$\mathcal{M}_q \Rightarrow |p(v, k) - 2^{-m}p(v)| \le \Delta_{s,k} + \Delta_k$$

$$\le (1 + q)p(v)2^{-m}\eta, \tag{37}$$

which is the event $\mathcal{N}_{\eta + q\eta}$. Now, we do the probabilistic part. We will show that $\mathcal{S}'$ probabilistically implies $\mathcal{M}_q$. We must compute $\Pr(\mathcal{S}' \wedge \bar{\mathcal{M}}_q)$. We will use $\Pr(\mathcal{S}' \wedge \bar{\mathcal{M}}_q) = \Pr(\mathcal{S}') - \Pr(\mathcal{S}' \wedge \mathcal{M}_q)$. So we will compute $\Pr(\mathcal{S}' \wedge \mathcal{M}_q)$. For every $v$ fixed, let $M_q^{(0)}(v)$ and $M_q^{(1)}(v)$ respectively denote the set of values $k$ such that $\mathcal{M}_q^{(0)}(k, v)$ and $\mathcal{M}_q^{(1)}(k, v)$ are TRUE. We obtain

$$M_q^{(0)}(v) = \{k | a_0(k, v) \le q\gamma p(v)2^{-m}\},$$

and

$$M_q^{(1)}(v) = \{k | a_1(k, v) \le qp(v)2^{-m}\}.$$

Let $M_q(v) = M_q^{(0)}(v) \cap M_q^{(1)}(v)$. We will use Proposition 5 to obtain a lower bound on the size of these sets for every $v$ for which $\mathcal{S}'$ is TRUE. It is not too hard to see that $\Sigma_k 2^{-m}a_1(k, v) = p(v)$. For $v$ fixed, using Proposition 5 with $y = k \in \{0, 1\}m$, $a_y = a_1(k, v)$, $p(y) = 2^{-m}$, and $\mu = p(v)$, we obtain that $2^{-m}|M_q^{(1)}(v)| \ge (1 - 1/q)$. Similarly, $\Sigma_k 2^{-m}a_0(k, v) = \gamma p(v)$. For $v$ fixed, using Proposition 5 with $y = k$, $a_y = a_0(k, v)$, $p(y) = 2^{-m}$ and $\mu = \gamma p(v)$ we obtain that $2^{-m}|M_q^{(0)}(v)| \ge (1 - 1/q)$. Therefore, we have

$$|M_q(v)| \ge 2^m\left(1 - \frac{2}{q}\right). \tag{38}$$

Using (37) for the second inequality and (38) for the third inequality, we obtain

---

[4] In a previous version of the proof, the bounds on $a_0$ and $a_1$ included a large factor $2^m$. For $m$ fixed, this factor was a constant, but it was nevertheless annoying. This large factor was taken out with the precious help of Hitoshi Inamori.

$$\Pr(\mathcal{S}' \wedge \mathcal{M}_q)$$
$$\geq \sum_{v \in \mathcal{S}'} \sum_{k \in M_q(v)} p(k, v)$$
$$\geq \sum_{v \in \mathcal{S}'} p(v) \sum_{k \in M_q(v)} 2^{-m}(1 - (1 + q)\eta)$$
$$\geq \sum_{v \in \mathcal{S}'} p(v) 2^m \left(1 - \frac{2}{q}\right) 2^{-m}(1 - (1 + q)\eta)$$
$$\geq \sum_{v \in \mathcal{S}'} p(v) \left(1 - \frac{2}{q} - (1 + q)\eta\right)$$
$$\geq \Pr(\mathcal{S}') - \frac{2}{q} - (1 + q)\eta.$$

We obtain

$$\Pr(\mathcal{S}' \wedge \bar{\mathcal{M}}_q) = \Pr(\mathcal{S}') - \Pr(\mathcal{S}' \wedge \mathcal{M}_q) \leq \frac{2}{q} + (1 + q)\eta.$$

If we set $q = \sqrt{2/\eta}$, we obtain

$$\Pr(\mathcal{S}' \wedge \bar{\mathcal{M}}_q) \leq \eta + 2\sqrt{2\eta}$$

or equivalently $\mathcal{S}' \Rightarrow_{\eta + 2\sqrt{2\eta}} \mathcal{M}_q$. Now, we recall that $\mathcal{M}_q$ with $q = \sqrt{2/\eta}$ implies the event $\mathcal{N}_\sigma$. We have thus shown $\mathcal{S}' \Rightarrow_{\eta + 2\sqrt{2\eta}} \mathcal{N}_\sigma$. This concludes the proof. □

One obtains Theorem 1 by combining the two probabilistic implications (i.e., Lemma 6 and Lemma 7) and Lemma 1.

## 6. *Conclusion*

The techniques that we have described here, some of them taken in Yao [1995], were proven to be efficient to analyze the security of quantum key distribution. However, these techniques were first used in Mayers [1996] to analyze a quantum protocol for a different application, a quantum string oblivious transfer protocol [Bennett et al. 1992]. For some time this quantum string oblivious transfer was ignored because it was built on top of a task called bit commitment. This was proven to be unsecure given that the participants, potential cheaters, have unlimited computational power [Mayers 1997]. However, recently a quantum protocol was proposed [Dumais et al. 2000] for bit commitment under some computational assumption and this raises the important question of the security of the quantum string oblivious transfer protocol on top of a computationally secure quantum bit commitment. We hope that the technique described here would be useful to address this question.

There is also the serious issue of defective and unreliable quantum apparatus. A more practical protocol were the encoding must still respect the exact polarization angle specified in the protocol, but not necessarily for a single photon, was proven secure in Inamori et al. [1999] using the techniques described here. The most powerful and global approach to address this problem is proposed in Mayers and Yao [1998] and Mayers [2001a]. However, the results in Mayers and Yao [1998] and Mayers [2001a] are general and their applicability to a given protocol is still an open question. Again, we hope that the techniques provided here will be useful to establish the connection.

*Note added*. An alternative proof for the security of the BB84 protocol was proposed by Shor and Preskill [2000]. This last security result is weaker than the result here because their proof requires the assumption that Bob's measuring apparatus is perfect (or close to perfect eventually). This assumption can certainly help to simplify the proof, but it is a step backward with respect to the ultimate objective, which is to trust only restricted and simple properties of the apparatus used. Here, we need to trust only a very natural property of Bob's measuring apparatus, and the problem of the untrusted source is taken care in Mayers and Yao [1998] and Mayers [2001a].

## Appendixes

### Appendix A Notations

#### A.1. Integers and Integer-Valued Function

$N$:  The number of photons sent in the protocol.

$m$:  The length of the key.

$n_E$:  The integer $|E|$. The length of the raw key.

$d_X(g, h)$:  The Hamming distance between $g$ and $h$ on $X$, that is, $d_X(g, h) = |\{i \in X | g[i] \neq h[i]\}|$.

$d_+(y)$:  An integer valued function that is convenient in the proof. For every $y > 0$, $d_+(y) = \lceil (\delta + y)p_E n_\Omega \rceil$.

$d$:  The number $\lceil \delta p_T n_\Omega \rceil$. It is the number of errors tolerated in the test executed on $T$.

$d'$:  The number $d_+(\beta)$. The number of errors on $E$ that is tolerated by the error-correction code.

$d_T$:  The number of errors that actually occur on $T$.

$d_E$:  The number of errors that actually occur on $E$.

$r$:  The number of bits for error-correction, that is, the length of the syndrome.

$d_W$:  The minimal weight of the strings in $G^*$.

#### A.2. Real Numbers

$p_T$:  The probability that a position $i \in \Omega$ is tested (i.e., included in $R$).

$\delta$:  A parameter in the protocol that determines the number of errors $d$ tolerated in the test via the formula $d = \delta p_T n_\Omega$. Normally, $\delta$ should be slightly greater than the expected error rate in the quantum channel.

$\beta$:  A parameter in the protocol that uniquely determines $d'$, the number of errors tolerated by the error-correcting code.

$\epsilon$:  A parameter similar to $\beta$ but used to determine $d_+(\epsilon)$, the number of errors tolerated in a fictive test executed on $E$.

$\mu(\beta, n_\Omega)$:    Exponentially small number that corresponds to the probability of $d_E \geq d_+(\beta)$ where $d_E$ is the number of errors in $E$ and $p_E = 1 - p_T$ is the probability that a position $i$ is not tested.

$\lambda$:    The real number $2^{-\tau n_E}$. It is an exponentially small upper bound on the probability that $d_W < H^{-1}\left(1 - (m + r)/n_E - \tau\right) \times n_E$.

$\gamma$:    The real number $\mu(\epsilon, n_\Omega)^{1/2}$. It is an exponentially small number that occurs in the proof of privacy because of Proposition 5.

$\eta$:    The sum $2\sqrt{\gamma} + \gamma$.

$\xi$:    The sum $\gamma + \lambda + \eta + 2\sqrt{2\eta}$. It is an exponentially small parameter used (together with $\sigma$) to indicate the level of privacy.

$\sigma$:    The exponentially small number $\eta + \sqrt{2\eta}$ used (together with $\xi$) to indicate the level of privacy.

### A.3. *Strings and String-Valued Function*

$a$:    Alice's string of bases.

$g$:    Alice's string of bits.

$b$:    Bob's string of bases.

$h$:    Bob's string of bits (outcomes of measurements).

$\bar{g}$:    The substring $g[E]$ of $g$. It is called the raw key.

$s$:    The string $F \bullet \bar{g}$, where $\bar{g} = g[E]$. It is called the syndrome.

$k$:    The (final) key shared by Alice and Bob.

$\bar{b}$:    The string of bases used by Bob in the modified protocol. The bases used for the positions outside $R$ are flipped.

$\lambda$:    The string valued function that on any $\alpha \in C^\perp[G]$ returns the unique string $\lambda(\alpha)$ such that $\lambda(\alpha) \bullet G = (\alpha)$.

$\hat{K}$:    The random bits in Alice's random tape that will be used to generate the matrix $K$.

### A.4. *Composites Values*

Formally, any value can be represented by a string over some alphabet. It is only with respect to the description of the protocol that the following values are said to be "composite."

$\hat{c}$:    The content of the random tape initialized at the beginning of the protocol. Each participant has a part of $\hat{c}$. In the modified protocol, $\hat{c} = (\bar{b}, a, R, g, \hat{K})$.

$\hat{q}$:    The outcome of the overall measurement executed jointly by all participants in view of $\hat{c}$. In the modified protocol, $\hat{q} = (\mathcal{D}, h, j)$.

$\hat{v}$:    The overall classical outcome of the protocol. It includes both $\hat{c}$ and $\hat{q}$: $\hat{v} = (\hat{c}, \hat{q})$.

$c$:    The classical information received by Eve (i.e., Eve–Bob in the modified protocol) and that is not the direct result of a quantum measurement. It is also a function of $\hat{v}$, not necessarily of $\hat{c}$ alone. In the modified protocol, $c = (\bar{b}, a, R, g[\bar{E}], K, s)$.

$j$:    The outcome of the final measurement executed by Eve–Bob.

$q$:    The quantum outcome received by Eve (i.e., Eve–Bob in the modified protocol). In the modified protocol, $q = (\mathcal{D}, h, j)$.

$v$:     Eve's view, that is, all classical data received by Eve including outcomes of measurements: $v = (c, q)$. It is a deterministic function of $\hat{v}$. The same notation is used in the original as in the modified protocol, but in the modified protocol it is called Eve–Bob's view.

$z$:     The partial view $(\bar{b}, \mathcal{D}, h, a, R)$. It is a part of Eve–Bob's view that is important in the proof because it contains no information about the string $g$ and yet contains enough information to uniquely determine $\hat{\Pi}_0$ and $\hat{\Pi}_1$.

### A.5. Set of Strings and Binary Matrix

$G^*$:   The set of linear combinations of rows in $F$ or $K$ with at the least one row in $K$.

$C_s$:   The set of codewords $\{\alpha \in \{0, 1\}^E | F \bullet \alpha = s\}$. It is the set of codewords $\alpha \in \{0, 1\}^E$ consistent with the syndrome $s$.

$F$:     A $r \times n_E$ matrix used to define the syndrome $s$.

$K$:     A $m \times n_E$ matrix used to define the key $k$ via $k = K \bullet \bar{g}$, where $\bar{g} = g[E]$.

### A.6. Set of Positions

$\mathcal{D}$:   The set of positions where a photon is detected by Bob.

$\Omega$:   The set of positions $i \in \mathcal{D}$ such that $a[i] = b[i]$.

$R$:     Random set of positions used for testing. Every $i = 1, \ldots, N$ is put in $R$ with probability $p_T$.

$T$:     The set $\Omega \cap R$. It is the set of tested positions.

$E$:     The set $\Omega - R$. It is the set of positions used to define the raw key $\bar{g} = g[E]$.

### A.7. Events

$N_\sigma$:   The event that the view $v$ is $\sigma$-informative about the key $k$: $|p(k|v) - (1/2^m)| \leq \sigma/2^m$.

$\mathscr{P}_T$:   The event $d_T \geq \delta p_T n_\Omega$. It is the event that is TRUE when the test on $T$ passes.

$\mathscr{P}_E$:   The event $d_E \geq \delta p_E n_\Omega$. It is the event that is TRUE when the fictive test on $E$ passes.

$\mathscr{X}_1$:   The validation constraint $d' \geq d_+(\beta)$ that is required by the error correction procedure.

$\mathscr{X}_2$:   The validation constraint $d_W \geq 2d_+(\epsilon)$. This constraint satisfies the hypothesis of Lemma 5, but unfortunately it is a hard problem to check if this constraint holds.

$\mathscr{X}'_2$:   The validation constraint $H^{-1}(1 - (r + m)/n_E - \tau)n_E \geq 2d_+(\epsilon)$, where $\epsilon > 0$ and $\tau > 0$ are any positive values fixed in the protocol. It is an alternative to $\mathscr{X}_2$ because $d_W/2 \geq H^{-1}(1 - (r + m)/n_E - \tau)n_E$ can be obtained probabilistically.

$\mathscr{X}_3$:   The validation constraint $n_E \geq n_E^{\min}$, $n_\Omega \geq n_\Omega^{\min}$ and $m \leq m^{\max}$, which is necessary so that the bound on Eve's information is a fixed number, not a random number.

### A.8. States and Density Matrices

$\Psi(g, a)$:   For any string of bits $g$ and string of bases $a$, $\Psi(g, a)$ is the BB84 encoding of the string $g$ in the bases $a$.

$\Psi_{\bar{E}}$: The state $\Psi_{\bar{E}} = \Psi(g[\bar{E}], a[\bar{E}])$ for the photons prepared by Alice in $\bar{E}$, where $\bar{E} = 1, \ldots, N - E$.

$|\phi v\rangle$: The state so that $\langle \phi_{c,q}|$ is the collapse operation on the photons associated with the view $v = (c, q)$. We have $|\phi_{c,q}\rangle \langle \phi_{c,q}| \stackrel{def}{=} E_{q|c}$. (See definition of the collapse operation $E_{q|c}$.)

$\bar{\phi}_{c,q}$: The state $\langle \Psi_{\bar{E}}|\phi_{c,q}\rangle$ for the photons in $E$.

$\rho_a$: Density matrix for the original random state $\Psi(g, a)$ with $a$ fixed. It's the fully mixed density matrix $2^{-N}\mathbf{I}$.

$\rho_{\bar{b}}$: Density matrix similar to $\rho_a$ except that it is for the random state $\Psi(g, \bar{b})$. It's also the fully mixed density matrix $2^{-N}\mathbf{I}$.

$\rho_{c|q}$: It is the density matrix for all the photons given that the classical information $c$ is known. The dependence on $q$ is related to the fact that the function $\mathbf{c}$ on the overall classical random tape $\hat{c}$ used in the protocol might itself depend on the outcome of a quantum measurement.

$\bar{\rho}_s$: The matrix $|C_s|^{-1} \sum_{\alpha \in C_s} |\Psi(\alpha, E)\rangle\langle\Psi(\alpha, E)|$. It is the density matrix of the photons in $E$ given that the syndrome $s$ is known.

$\bar{\rho}_{s,k}$: A density matrix for the photons in $E$ like $\bar{\rho}_s$ but given that $s$ and $k$ is known.

## A.9. *Operators and Projections*

$E_{q|c}$: It is the positive operator on the state space $H^Q$ of the photons associated with the measurement outcome $q$ executed in view of $c$.

$P_{c|q}$: The projection operator on the classical part $H^C$ associated with the classical announcement $c$ in the view $v = (c, q)$.

$E_v = E_{(c,q)}$: The operator $P_{c|q} \otimes E_{q|c}$ associated with the view $v = (c, q)$. It acts on the state space $H^C \otimes H^Q$ where $H^C$ is the classical part and $H^Q = H^A$ is the quantum part.

$\Pi_0[X, d]$ **and** $\Pi_1[X, d]$: For any set of positions $X$ and integer $d > 0$, $\Pi_0[X, d]$ is the projection on the span of $\{\Psi(g, b)|d_X(g, h) \geq d\}$, and $\Pi_1[X, d] = \mathbf{I} - \Pi_0[X, d]$.

$\hat{\Pi}_0$: The projection $\Pi_0[E, d_+(\epsilon)]$.

$\hat{\Pi}_1$: The projection $\Pi_1[T, d]$.

$\bar{\Pi}_0$: The projection $\Pi_0[E, \lfloor d_W/2 \rfloor]$.

$\bar{\Pi}_1$: The projection $\mathbf{I} - \bar{\Pi}_0$.

## *Appendix B. Mutual Information*

Privacy is often expressed in terms of *mutual information* or *Shannon's entropy*. It is impossible to do justice in one small subsection to the concepts of mutual information and Shannon's entropy. Here some simple techniques and formulas are listed. Let $X$, $Y$, and $Z$ be any three random variables. Let $p(x, y) = \Pr(X = x \wedge Y = y)$, $p(x) = \Pr(X = x)$, $p(y) = \Pr(Y = y)$, $p(x|y) = \Pr(X = x|Y = y)$, $p(x, y, z) = \Pr(X = x \wedge Y = y \wedge Z = z)$, etc.

*Definition* 7.  The mutual information between $X$ and $Y$ is given by

$$I(X;Y) = \sum_{x,y} p(x, y) \log_2 \left( \frac{p(x,y)}{p(x)p(y)} \right).$$

*Definition* 8.  The Shannon entropy of $X$ is given by

$$H(X) = I(X;X) = - \sum_{x} p(x) \log_2 p(x).$$

*Definition* 9.  The conditional Shannon entropy of $X$ given $Y$ is given by

$$H(X|Y) = - \sum_{x,y} p(x, y) \log_2 p(x|y).$$

One can easily verify that $I(X;Y) = H(X) - H(X|Y)$ as follows: Note that $I(X;Y)$ is the expected value of $\log_2 (p(x, y)/p(x)p(y))$:

$$I(X;Y) = E \left( \log_2 \left( \frac{p(x,y)}{p(x)p(y)} \right) \right).$$

Similarly, one has

$$H(X) = E(-\log_2 p(x)) \qquad \text{and} \qquad H(X|Y) = E(-\log_2 p(x|y)).$$

Thus, one obtains

$$H(X) - H(X|Y) = E(-\log_2 p(x)) - E(-\log_2 p(x|y))$$
$$= E(-\log_2 p(x) + \log_2 p(x|y))$$
$$= E \left( \log_2 \frac{p(x,y)}{(p(x)p(y))} \right) = I(X;Y).$$

By symmetry one has also $I(X;Y) = H(Y) - H(Y|X)$.

*Definition* 10.  The conditional mutual information between $X$ and $Y$ given an event $\mathscr{E}$ is

$$I(X;Y|\mathscr{E}) = \sum_{x,y,z} p(x, y|\mathscr{E}) \log_2 \left( \frac{p(x,y|\mathscr{E})}{p(x|\mathscr{E})p(y|\mathscr{E})} \right).$$

*Definition* 11.  The conditional mutual information between $X$ and $Y$ given $Z$ is

$$I(X;Y|Z) \overset{def}{=} \sum_{x,y,z} p(x, y, z) \log_2 \left( \frac{p(x,y|z)}{p(x|z)p(y|z)} \right)$$
$$= \sum_{z} p(z) I(X;Y|Z = z).$$

One can verify that

$$I(X;Y, Z) = I(X;Y|Z) + I(X;Z). \tag{39}$$

Many other formulas of the same kind can be obtained. For instance,

$$I(X;Y) = I(X;Y|Z) + I(X;Z) + I(Y;Z) - I(X, Y;Z),$$

because

$$\frac{p(x, y)}{p(x)p(y)} = \frac{p(x, y|z)}{p(x|z)p(y|z)} \times \frac{p(x, z)}{p(x)p(z)}$$
$$\times \frac{p(y, z)}{p(y)p(z)} \times \frac{p(x, y)p(z)}{p(x, y, z)}.$$

## *Appendix C. Linear Codes*

In this appendix, we give some minimal about binary linear codes.

### C.1. *Binary Strings and Matrices*

A *binary string* $x$ of length $n$ is a mapping from $\{1, \ldots, n\}$ into $\{0, 1\}$ or alternatively an $n$-tuple in $\{0, 1\}^n$. The length of $x$ is denoted by $|x|$. The *weight* of $x$, that is, the number of 1 in $x$, is denoted by $\#(x)$. We define the *minimal weight* of a set $A \subseteq \{0, 1\}^{n_E}$ as the minimum of $\#(w)$ over all $w \in A$. The $i$th element of $x$ is $x[i]$; the sum $x \oplus y$ where $x, y \in \{0, 1\}^n$ is given by

$$(x \oplus y)[i] \stackrel{def}{=} x[i] \oplus y[i] \stackrel{def}{=} \begin{cases} 0 & \text{if} \quad x[i] = y[i] \\ 1 & \text{if} \quad x[i] \neq y[i]. \end{cases}$$

A sum of two or more strings is also called a *linear combination*. If $A$ and $B$ are two sets of strings in $\{0, 1\}^n$, we define $A \oplus B = \{w | w = w_1 \oplus w_2 \text{ where } w_1 \in A \text{ and } w_2 \in B\}$. The inner product of $x$ and $y$ is $x \bullet y = \oplus_{j=1}^n x[j]y[j]$.

For every set $E \subseteq \{1, \ldots, n\}$, we denote by $\{0, 1\}^E$ the set of mappings from $E$ to $\{0, 1\}$. These mappings are also called binary strings. For every subset $E \subseteq \{1, \ldots, n\}$, we use $x[E]$ to denote the substring of $x$ restricted to $E$. The substring $x[E]$ is the unique mapping $\bar{x}$ from $E$ to $\{0, 1\}$ such that $\bar{x}[i] = x[i]$ for all $i \in E$.

An $n_1 \times n_2$ *binary matrix* $M$ is a mapping from $\{1, \ldots, n_1\} \times \{1, \ldots, n_2\}$ into $\{0, 1\}$. The $(i, j)$th element of a matrix $M$ is denoted as $M[i, j]$. A $r \times r$ square matrix that contains 1 everywhere in the diagonal and 0 elsewhere is denoted $\mathbf{I}_r$. A $r \times k$ matrix which contains 0 everywhere is denoted $\mathbf{0}_{r,k}$. The $i$th *row* and the $j$th *column* of $M$ are strings noted $M[i, \cdot]$ and $M[\cdot, j]$ respectively. The transpose $M^T$ of the $n_1 \times n_2$ matrix $M$ is the $n_2 \times n_1$ matrix given by $M^T[i, j] = M[j, i]$. If $M$ is an $n_1 \times n_2$ binary matrix and $x$ is a binary string of length $n_2$, the product $M \bullet x$ is the string of length $n_1$ given by

$$M \bullet x[i] \stackrel{def}{=} \bigoplus_{j=1}^{n_2} M[i, j]x[j].$$

We could also write $M \bullet x = \oplus_j x[j]M[\cdot, j]$ (i.e., as the linear combination of the columns $M[\cdot, j]$, where $x[j] = 1$). If $x$ is a string of length $n_1$, then $x \bullet M$ is the string of length $n_2$ given by

$$x \bullet M[j] \overset{def}{=} \overset{n_1}{\underset{i=1}{\oplus}} x[i]M[i,j].$$

We could also write $(x \bullet M) = \oplus_i x[i]M[i, \cdot]$ (i.e., as the linear combination of the rows $M[i, \cdot]$ where $x[i] = 1$). Note that the string $x \bullet M^T$ and the string $M \bullet x$ are exactly the same string. The product of a $k \times n$ matrix $G$ with a $n \times r$ matrix $F$ is the $k \times r$ matrix $G \bullet H$ given by

$$G \bullet H[i, j] \overset{def}{=} \overset{n}{\underset{z=1}{\oplus}} G[i, z]H[z, j] \overset{def}{=} G[i, \cdot] \bullet H[\cdot, j].$$

The strings $x_1, \ldots, x_r$ are *linearly independent* if no linear combination over a subset of these strings is the string $\mathbf{0}$. One can easily check that the rows of a $r \times n$ binary matrix $M$ are linearly independent if and only if, for any two distinct strings $x$, $y$ of length $r$, $x \bullet M \neq y \bullet M$. So, if the rows of $M$ are linearly independent, there are $2^r$ distinct linear combinations of this kind.

Let $E \subseteq \{1, 2, \cdots, n\}$. For every pair of strings $\alpha$, $\alpha' \in E^{\{0,1\}}$, for every subset $X \subseteq E$, we define $d_X(\alpha, \alpha') = \#(\alpha[X] \oplus \alpha'[X]) = |\{i \in X | \alpha[i] \neq \alpha'[i]\}|$. If $X = E = \{1, \ldots, r\}$, then $d_X(\alpha, \alpha')$ is the usual Hamming distance on strings of length $r$. The following lemma is a useful tool.

LEMMA 8. *Let $S_d$ be the set of strings $w \in \{0, 1\}^n$ with $\#(w) \leq d$. Let $p = d/n$ and $q = 1 - p$. If $d < n/2$, then*

$$\frac{2^{H(p) \times n}}{\sqrt{8pqn}} \leq |S_d| \leq 2^{H(p) \times n},$$

*where $H(p) = -(p \log_2 p + q \log_2 q)$.*

Lemma 8 follows from the standard bounds on binomial coefficients [MacWilliams and Sloane 1977].

## C.2. *Error-Correcting Code*

Let $1 \leq r \leq n$ and $k = n - r$. Consider any $r \times n$ binary matrix $F$ with $r$ linearly independent rows. The set $C[F] = \{w \in \{0, 1\}^n | F \bullet w = \mathbf{0}\}$ is called a $(n, k)$-*linear code*. There are $2^k$ codewords in $C[F]$. The matrix $F$ is called the *parity-check matrix* of the $(n, k)$-linear code $C[F]$. For every linear code $C$ and every string $w$ of length $n$, the set $C \oplus w = \{u \oplus w | u \in C\}$ is a *coset* of $C$. One can easily check that, for every string $s$ of length $r$, the set $\{w \in \{0, 1\}^n | F \bullet w = s\}$ is a coset of $C[F]$; we denote this coset by $C[F, s]$. The string $s = F \bullet w$ is called the *syndrome* of $w$ associated with $F$.

For any $(n, k)$-linear code $C$, the *dual* of $C$, denoted as $C^\perp$, is the set of strings $x$ such that $(\forall w \in C) \ w \bullet x = 0$. The dual $C^\perp$ is a $(n, r)$-linear code. We denote by $C^\perp[F] = \{w \in \{0, 1\}^n | w = \lambda \bullet F, \lambda \in \{0, 1\}^r\}$ the set of linear combinations of rows $F[i, \cdot]$ in the $r \times n$ matrix $F$. One can easily check that the dual of $C[F]$ is the set of linear combinations of rows in $F$, that is, $C[F]^{\perp} = C^{\perp}[F]$. The matrix $F$ is called a *generator* matrix of the $(n, r)$-linear code $C^{\perp}[F]$. So the $r \times n$ matrix $F$ is both a parity-check matrix for the $(k, n)$-linear code $C[F]$ and a generator matrix for the $(n, r)$-linear code $C^{\perp}[F]$. There are $2^r$ codewords in $C^{\perp}[F]$. One can also construct a matrix $G$ that is both a parity check matrix for $C^{\perp}[F]$ and a generator matrix for $C[F]$; any $k \times n$ matrix $G$ that contains $k$ independent rows in $C[F]$ will suffice.

An $(n, k)$-linear error-correcting code $C$ usually comes with an encoding procedure *Enc* that maps a message $x \in \{0, 1\}^k$ into a codeword $w = Enc(x) \in C$. Of course, the mapping defined by *Enc* must be one-to-one otherwise some information about $w$ would be lost. The codeword $w$ is sent into some channel and a string $w'$ is obtained on the other side. Then usually an error-correcting procedure is executed to map the string $w' \in \{0, 1\}^n$ into a codeword in $C$. If the number of errors in $w'$ is sufficiently small then, with probability almost 1 (if not 1), this codeword is the original codeword $w$. Then the codeword $w$ is mapped back into the original message $x$.

*Appendix D. The Density Matrices $\tilde{\rho}$*

Consider a linear code $C[G] \subseteq \{0, 1\}^n$ of dimension $q$ and a coset $C[G, x]$ of this code ($G$ is the parity check matrix and $x$ is the syndrome). Here, we analyze the general situation where a string $\tilde{g}$ uniformly chosen at random in the coset $C[G, x]$ is sent from Alice to Bob using a fixed string of bases $a \in \{+, \times\}^n$. We want to find the matrix representation of the density operator

$$\tilde{\rho}_x = 2^{-q} \sum_{\tilde{g} \in C[G,x]} \tilde{\Psi}(\tilde{g}, a)\tilde{\Psi}(\tilde{g}, a)^{\dagger}$$

in the basis $\{\tilde{\Psi}(\alpha, b) | \alpha \in \{0, 1\}^n\}$, where $b = \bar{a}$. To apply this result to this paper, one must use

$$G = \begin{pmatrix} F \\ K \end{pmatrix},$$

$x = (s, k)$ and $q = n - r - m$ but the computation for the general case is the same. A key ingredient is that if a string $g$ belongs to a code $C = C[G]$ for which $G$ is the parity check matrix then we have $g = \lambda \bullet G^{\perp}$ where $\lambda \in \{0, 1\}^{\dim C}$ and $G^{\perp}$ is a parity check matrix for the dual code. We will apply this principle twice, once with the code and once with its dual. We have

$$\tilde{\rho}_x = \frac{1}{|C|} \sum_{g \in C[G,x]} |w\rangle\langle w|.$$

We will use the fact that in the conjugate basis we have

$$|w\rangle = 2^{-n} \sum_{t \in \{0,1\}^n} (-1)^{g \bullet t} |t\rangle.$$

We obtain

$$\tilde{\rho}_x = \frac{2^{-n}}{|C|} \sum_{t,t',g \in C} (-1)^{g \bullet (t \oplus t')} |t\rangle\langle t'|.$$

Let $g_0$ be any string in the coset $C[G, x]$. We will use the fact that the sum over $g \in C[G, x]$ can be replaced by a sum over $\gamma \in \{0, 1\}^{\dim C}$ with the change of variable $g \mapsto (\gamma \bullet G^{\perp}) \oplus g_0$. We get

$$\tilde{\rho}_x = \frac{2^{-n}}{|C|} \sum_{t,t',\gamma \in \{0,1\}^{\dim C}} (-1)^{(g_0 \oplus \gamma \bullet G^{\perp}) \bullet (t \oplus t')} |t\rangle\langle t'|.$$

After simple algebra, we get

$$\tilde{\rho}_x = \frac{2^{-n}}{|C|} \sum_{t,t'} (-1)^{g_0 \bullet (t \oplus t')} \underbrace{\sum_{\gamma \in \{0,1\}^{\dim C}} (-1)^{\gamma \bullet G^{\perp} \bullet (t \oplus t')}}_{k(t,t')} |t\rangle\langle t'|.$$

Now, consider the coefficient $k(t, t')$. This coefficient vanishes if $G^{\perp} \bullet (t \oplus t') \neq 0$, that is, if $(t \oplus t') \notin C^{\perp}$. If $(t \oplus t') \in C^{\perp}$, we have $k(t, t') = |C|$. We obtain

$$\tilde{\rho}_x = 2^{-n} \sum_{t,t'|(t \oplus t') \in C^{\perp}} (-1)^{g_0 \bullet (t \oplus t')} |t\rangle\langle t'|,$$

where we used $g_0 \bullet (t \oplus t') = (t \oplus t') \bullet g_0$. Now, we will use the fact that $(t \oplus t')$ is a string in $C^{\perp}$. We obtain that $t \oplus t' = \lambda(t \oplus t') \bullet G$, where $\lambda(t \oplus t')$ is the unique string with this property. The exponent $(t \oplus t') \bullet g_0$ becomes $\lambda(t \oplus t') \bullet G \bullet g_0 = \lambda(t \oplus t') \bullet x$, by definition of $g_0$. We obtain

$$\tilde{\rho}_x = 2^{-n} \sum_{t,t'|(t \oplus t') \in C^{\perp}} (-1)^{\lambda(t \oplus t') \bullet x} |t\rangle\langle t'|.$$

or equivalently

$$\langle t|\tilde{\rho}_x|t'\rangle = 2^{-n} \begin{cases} (-1)^{\lambda(t \oplus t') \bullet x} & \text{if} \quad (t \oplus t') \in C^{\perp} \\ 0 & \text{otherwise.} \end{cases}$$

### D.1. *An Alternative Computation*

Here is an alternative computation (the one that was used in the original proof [Mayers 1966]). We need some definitions. For every vector $\theta \in \{0, 1\}^n$, let us define a unitary transformation $U_\theta$ on the state space of the photons:

$$U_\theta \tilde{\Psi}(\tilde{g}, a) = \tilde{\Psi}(\theta \oplus \tilde{g}, a).$$

One can easily check that $U_\theta$ is in fact a product of unitary mappings $U_\theta = U_{\theta[1]} \cdots U_{\theta[n]}$ where $U_{\theta[i]}$ is defined on the state space for the $i$th photon. For every position $i$ where $\theta_i = 1$, the transformation $U_{\theta[i]}$ maps the state $\tilde{\Psi}_{[i]}(0, b[i])$ into itself and the state $\tilde{\Psi}_{[i]}(1, b[i])$ into $-\tilde{\Psi}_{[i]}(1, b[i])$. So, if there is an even number of positions $i$ where $\alpha_i = \theta_i = 1$, one has

$$U_\theta \tilde{\Psi}(\alpha, b) = \tilde{\Psi}(\alpha, b);$$

otherwise, one has

$$U_\theta \bar{\Psi}(\alpha, b) = -\bar{\Psi}(\alpha, b).$$

In terms of the inner product $\bullet$ on the vector space $\{0, 1\}^n$, one has

$$U_\theta \bar{\Psi}(\alpha, b) = \begin{cases} \bar{\Psi}(\alpha, b) & \text{if} \quad \theta \bullet \alpha = 0 \\ -\bar{\Psi}(\alpha, b) & \text{if} \quad \theta \bullet \alpha = 1. \end{cases}$$

For every $\theta \in C[G, x]$, one has $C[G, x] = C[G, \mathbf{0}] \oplus \theta$. Therefore, for every $\theta \in C[G, x]$,

$$\hat{\rho}_x = U_\theta \hat{\rho}_0 U_\theta, \tag{40}$$

where $U_\theta^\dagger = U_\theta$ was used. For any operator $\hat{\rho}$ and any $\theta$, one may easily check that, in the basis $\{\bar{\Psi}(\alpha, b) | \alpha \in \{0, 1\}^n\}$,

$$
\begin{aligned}
(U_\theta \hat{\rho} U_\theta)_{\alpha, \alpha'} &\overset{\text{def}}{=} \bar{\Psi}(\alpha, b)^\dagger (U_\theta \hat{\rho} U_\theta) \bar{\Psi}(\alpha', b) \\
&= (-1)^{(\alpha \oplus \alpha') \bullet \theta} \times (\hat{\rho})_{\alpha, \alpha'}.
\end{aligned}
\tag{41}
$$

Therefore, in view of (40) and (41), it is sufficient to obtain the matrix representation of the density operator $\hat{\rho}_0$ in Bob's basis.

Let $q$ be dimension of $C[G]$ (in this paper $q = n - r - m$ is the dimension of the code $C[F] \cap C[K]$). Let $\{\theta_1, \ldots, \theta_q\}$ be $q$ independent strings in $C[G]$. For every $j = 1, \ldots, q$, let $G^{(j)}$ be the span of $\{\theta_1, \ldots, \theta_j\}$ and $\hat{\rho}^{(j)} = 2^{-j} \sum_{\theta \in G^{(j)}} \bar{\Psi}(\theta, a) \bar{\Psi}(\theta, a)^\dagger$. Note that $\hat{\rho}_0 = \hat{\rho}^{(q)}$ and $C[G] = G^{(q)}$. We will show by induction on $j$, that for $j = 0, \ldots, \bar{q}$,

$$(\hat{\rho}^{(j)})_{\alpha, \alpha'} = 2^{-n} \times \begin{cases} 1 & \text{if} \quad (\alpha \oplus \alpha') \in G^{(j)\perp} \\ 0 & \text{otherwise.} \end{cases} \tag{42}$$

The case $j = 0$ can be easily computed: $G^{(0)} = \{0\}$ and $G^{(0)\perp} = \{0, 1\}^n$. Let us assume that (42) holds for $j$ and obtain it for $j + 1$. Because $G^{(j+1)} = G^{(j)} \cup (G^{(j)} \oplus \theta_{j+1})$, one has that

$$\hat{\rho}^{(j+1)} = \frac{1}{2} (\hat{\rho}^{(j)} + U_{\theta_{j+1}} \hat{\rho}^{(j)} U_{\theta_{j+1}}). \tag{43}$$

Therefore, using formula (41), one obtains

$$(\hat{\rho}^{(j+1)})_{\alpha, \alpha'} = \frac{1}{2} (\hat{\rho}^{(j)})_{\alpha, \alpha'} (1 + (-1)^{(\alpha \oplus \alpha') \bullet \theta_{j+1}}).$$

Note that $(\hat{\rho}^{(j+1)})_{\alpha, \alpha'}$ is either 0 or $2^{-n}$. One obtains that $(\hat{\rho}^{(j+1)})_{\alpha, \alpha'} = 2^{-n}$ if and only if $(\hat{\rho}^{(j)})_{\alpha, \alpha'} \neq 0$ and $(\alpha \oplus \alpha') \bullet \theta_{j+1} = 0$. So, $(\hat{\rho}^{(j+1)})_{\alpha, \alpha'} = 2^{-n}$ if and only if, for every $\theta \in G^{(j+1)}$, $(\alpha \oplus \alpha') \bullet \theta = 0$. This last condition is equivalent to $(\alpha \oplus \alpha') \in G^{(j+1)\perp}$. This concludes the induction.

Now, using the formula for $\hat{\rho}_0 = \hat{\rho}^{(q)}$ given by (42), together with formula (40) and (41), one obtains that, for every $\theta \in C[G, x]$,

$$(\hat{\rho}_x)_{\alpha, \alpha'} = 2^{-n} \times \begin{cases} 0 & \text{if } (\alpha \oplus \alpha') \notin C^{\perp}[G] \\ (-1)^{(\alpha \oplus \alpha') \bullet \theta} & \text{otherwise.} \end{cases}$$

Now, because the rows in $G$ are independent, for every $\alpha \in C^{\perp}[G]$, there is a unique string $\lambda(\alpha) \in \{0, 1\}^{r+m}$ such that $\lambda(\alpha) \bullet G = \alpha$. Let $\lambda(\alpha \oplus \alpha')$ be such that

$$\lambda(\alpha \oplus \alpha') \bullet G = (\alpha \oplus \alpha').$$

By definition of $\theta$, one has that $G \bullet \theta = x$. So, one obtains

$$(\hat{\rho}_x)_{\alpha, \alpha'} = 2^{-n} \times \begin{cases} 0 & \text{if } (\alpha \oplus \alpha') \notin C^{\perp}[G] \\ (-1)^{\lambda(\alpha \oplus \alpha') \bullet x} & \text{otherwise.} \end{cases} \tag{44}$$

*Appendix E. On Probabilistic Implications*

In our proof, the number $n_D$ in the fictive test lemma is itself random, whereas we want a fixed upper bound on Eve's information. To address this issue, we could use the law of large number to obtain a lower bound $n_D^{\min}$ for $n_D$ and use this lower bound instead of $n_D$, but, as we now explain, this approach would somehow ignore an essential mechanism used in the protocol. If we used the law of large numbers, it would mean that we want a bound on $n_D$ that is respected with probability almost one. This approach would be justified if whenever this bound is not respected, Eve receives too much information. However, this is not the way the protocol works. In our protocol, Eve receives no information when the bound is not respected because $m$ is set to 0 when it happens. Note that, as far as privacy is concerned, this mechanism works even if the lower bounds are very large and fails most of the time. Therefore, we should not have to use the law of large number in this particular context to prove privacy. One should use the law of large numbers to prove that $m$ is larger than 0 with a reasonable probability, but this is a different issue that is not related to privacy. The following proposition explains the basic mechanism that is used in our protocol, as far as privacy is concerned, to address the issue of random parameters such as $n_D$.

For concreteness, one can think that this proposition is used with the events $\mathcal{N}(l) : |p(k|v) - 2^{-m}| \leq \sigma(l)$, $l = 1, 2, \ldots$, where $\sigma(l)$ is some nonincreasing function of the integer $l$. The event $|p(k|v) - 2^{-m}| \leq \sigma$ (for some $\sigma > 0$) is used in Lemma 1. With this particular choice for $\mathcal{N}(l)$, the conclusion in Proposition 6 is essentially the kind of hypothesis that is required in Lemma 1.

PROPOSITION 6. *Consider a fixed protocol* (*i.e.*, *consider that all parameters are fixed*). *Let* $\xi = \xi(l)$ *be a nonincreasing real valued function defined on the set of positive integers* $l$. *Let* $\mathcal{N}(1), \mathcal{N}(2), \ldots$ *be events that are* (*simultaneously*) *defined by the protocol. Assume furthermore that, if* $l' \geq l$, *we have* $\mathcal{N}(l') \Rightarrow \mathcal{N}(l)$. *Let* $\mathbf{n}$ *be a random integer defined in the protocol.* (*The letter* $n$ *is often used to denote a security parameter but here* $\mathbf{n}$ *is not a security parameter.*) *Consider some fixed integer* $n^{\min} > 0$. *Consider any event* $\mathcal{P}_T$ *such that we can prove* $\mathcal{P}_T \Rightarrow_{\xi(n)|\mathbf{n}=n} \mathcal{N}(n)$ *for every* $n > n^{\min}$ (*see Definition 2*). *Let* $\mathcal{X}$ *be the event which is* TRUE *when* $n \geq n^{\min}$. *We have* $\mathcal{P}_T \wedge \mathcal{X} \Rightarrow_{\xi(n^{\min})} \mathcal{N}(n^{\min})$.

*Remark.* We actually have a more complicated situation in our proof. We have two probabilistic implications: (1) $\mathscr{P}_T \Rightarrow_{\gamma(n_1)|\mathbf{n}_1=n_1} \mathscr{S}(n_1)$, for every $n_1$ and (2) $\mathscr{S}(n_1) \Rightarrow_{\lambda(n_2)|\mathbf{n}_2=n_2} \mathscr{N}(n_1, n_2)$, for every $n_1$, $n_2$, where $\mathscr{S}(n_1)$ is some intermediary event. We apply Proposition 6 on each implication separately (and thus avoid mutual conditioning of variables) and then use Proposition 1 to obtain the hypothesis of Lemma 1.

PROOF OF PROPOSITION 6. We first show $\mathscr{X} \wedge \mathscr{P}_T \Rightarrow_{\xi(n^{min})} \mathscr{N}(\mathbf{n})$, in which we do not yet consider the event $\mathscr{N}(n^{min})$, that is, we first show $\Pr(\mathscr{N}(\mathbf{n}) \wedge \mathscr{X} \wedge \mathscr{P}_T) \leq \xi(n^{min})$. The event $\mathscr{N}(\mathbf{n})$ is TRUE in a particular run of the protocol if $\mathbf{n}$ takes a value $n$ such that $\mathscr{N}(n)$ is TRUE. Therefore, we have

$$
\begin{aligned}
\Pr(\mathscr{N}&(\mathbf{n}) \wedge \mathscr{X} \wedge \mathscr{P}_T) \\
&= \sum_{n \geq n^{min}} \Pr(\mathscr{P}_T \wedge \mathscr{N}(n)|\mathbf{n} = n)\Pr(\mathbf{n} = n) \\
&\leq \sum_{n \geq n^{min}} \xi(n)\Pr(\mathbf{n} = n) \\
&\leq \xi(n^{min}) \sum_{n \geq n^{min}} \Pr(\mathbf{n} = n) \\
&= \xi(n^{min})\Pr(n \geq n^{min}) \leq \xi(n^{min}).
\end{aligned}
$$

So, using Proposition 1, we obtain $\mathscr{X} \wedge \mathscr{P}_T \Rightarrow_{\xi(n^{min})} \mathscr{N}(\mathbf{n}) \wedge \mathscr{X}$. We have proven the proposition since $\mathscr{N}(\mathbf{n}) \wedge \mathbf{n} \geq n^{min} \Rightarrow \mathscr{N}(n^{min})$. □

ACKNOWLEDGMENTS. The author is grateful to Charles Bennett, Howard E. Brandt, Gilles Brassard, Claude Crépeau, David Divincenzo, Hitoshi Inamori, Peter Shor, John Smolin, Alain Tapp, and Andrew Yao for helpful discussions and advice.